\documentclass[a4paper,fleqn,usenatbib]{mnras}

\usepackage{newtxtext,newtxmath}

\usepackage[T1]{fontenc}
\usepackage{ae,aecompl}

\usepackage{graphicx}	
\usepackage[]{subfig}
\usepackage{amsmath}	
\usepackage{amssymb}	
\usepackage{soul}		

\defcitealias{nusser2000least}{NB00}

\title[BAO reconstruction]
{BAO reconstruction: a swift numerical action method for massive spectroscopic surveys}

\author[E. Sarpa et al.]{
	E. Sarpa,$^{1,2}$\thanks{E-mail: elena.sarpa@lam.fr}
	C. Schimd,$^1$
	E. Branchini,$^{3,4,5}$
	S. Matarrese.$^{2,6,7,8}$
	\\ 
$^{1}$ Aix Marseille Univ, CNRS, LAM, Laboratoire d'Astrophysique de Marseille, Marseille, France \\
$^{2}$ Dipartimento di Fisica e Astronomia ``Galileo Galilei'', Universit\`a degli studi di Padova, Via F. Marzolo, 8, I-35131 Padova, Italy \\
$^{3}$ Dipartimento di Matematica e Fisica, Universit\`a degli studi Roma Tre, Via della Vasca Navale, 84, 00146 Roma, Italy \\
$^{4}$ INFN - Sezione di Roma Tre, via della Vasca Navale 84, I-00146 Roma, Italy\\
$^{5}$ INAF - Osservatorio Astronomico di Roma, via Frascati 33, I-00040 Monte Porzio Catone (RM), Italy\\
$^{6}$ INFN, Sezione di Padova, via F. Marzolo 8, I-35131, Padova, Italy\\
$^{7}$ INAF-Osservatorio Astronomico di Padova, Vicolo dell'Osservatorio 5, I-35122 Padova, Italy\\
$^{8}$ Gran Sasso Science Institute, Viale F. Crispi 7, I-67100 L'Aquila, Italy
}

\date{Accepted XXX. Received YYY; in original form ZZZ}

\pubyear{2018}

\begin{document}
\label{firstpage}
\pagerange{\pageref{firstpage}--\pageref{lastpage}}
\maketitle

\begin{abstract}
A new fully non-linear reconstruction algorithm for the accurate recovery of the Baryonic Acoustic Oscillations (BAO) scale in two-point correlation functions is proposed, based on the least-action principle and extending the Fast Action Minimisation method by \citet{nusser2000least}.
Especially designed for massive spectroscopic surveys, it is tested on dark-matter halo catalogues extracted from the \textsc{deus-fur} $\Lambda$CDM simulation to trace the trajectories of up to $\sim207,000$ haloes backward-in-time, well beyond the first-order Lagrangian approximation. The new algorithm successfully recovers the BAO feature in real and redshift-space in both the monopole and the anisotropic two-point correlation function, also for anomalous samples showing misplaced or absent signature of BAO. In redshift space, the non-linear displacement parameter $\Sigma_\mathrm{NL}$ is reduced from $11.8\pm0.3h^{-1}$Mpc at redshift $z=0$ to $4.0\pm0.5h^{-1}$Mpc at $z\simeq37$ after reconstruction.  A comparison with the first-order Lagrangian reconstruction is presented, showing that this techniques outperforms the linear approximation in recovering an unbiased measurement of the acoustic scale.
\end{abstract}





\section{Introduction}

Before recombination and on small scales, the acoustic oscillations of the primordial baryon-photon plasma propagate at relativistic speed driven by photon pressure. On large scales they are standing waves with the fundamental mode set by the sound horizon $r_\mathrm{s}$ \citep{PeeblesYu1970,SunyaevZeldovich1970} and overtones damped on scales $\lesssim 8 h^{-1}$Mpc because of the non-perfect coupling between baryons and photons \citep{Silk1968}. At recombination and afterward, while photons freely stream, the residual baryonic fluctuations play as additional seeds of clustering at the characteristic scale $r_\mathrm{s}^*\simeq 150$~Mpc, progressively driving an excess of clustering of the dominant, collisionless dark matter \citep{HuSugiyama1996,EisensteinHu1998,Eisenstein1998}.
On these scales the gravitational clustering is well described by the linear approximation until today. Moreover, $r_\mathrm{s}^*$ depends on the primordial baryon-to-photon ratio and on the matter density, extremely well constrained by the cosmic microwave background observations \citep{Planck2018cosmo}. The baryonic acoustic oscillations (BAO) scale $r_\mathrm{s}^*$ therefore provide a powerful and robust standard ruler to strongly constrain the expansion rate of the universe and dark energy \citep{BlakeGlazebrook2003,SeoEisenstein2003,HuHaiman2003,White2005}, though requiring large volumes to be convincingly measured \citep{Cole+2005,Eisenstein+2005}.

At low redshift the non-linear evolution of structures, the bias of tracers, and their peculiar velocities determine a mild degradation of the acoustic signature in the two-point correlation functions \citep{EisensteinSeoWhite2007,Mehta+2011}, that must be reduced in order to enhance the signal-to-noise ratio and achieve sub-percent precision cosmology goals. The resulting broadening and shift the acoustic peak in the correlation function, or equivalently the damping and phase-shift of the higher harmonics in the power spectrum, can indeed be alleviated by means of so-called reconstruction techniques designed to recover the initial, linear density field using a Lagrangian perturbative scheme  \citep{Eisenstein2007improving,PadmanabhanWhiteCohn2009}, which provides a good reconstruction also when applied to biased mass tracers such as galaxies \citep{NohWhitePadmanabhan2009}.
The most widely used technique \citep{Padmanabhan+2012} is based on the displacement of the tracers backward-in-time using the Zel'dovich approximation calculated from the local density field, in which a linear correction is applied to remove redshift-space distortion (RSD) that  actually occur on both small and large scales \citep{Kaiser1987}. Since its first application on the Sloan Digital Sky Survey (SDSS) DR7 LRG sample at redshift $z\simeq0.35$ \citep{Padmanabhan+2012}, this reconstruction technique has been routinely adopted for BAO studies based on the data releases DR9 to DR12 using both the LOWZ and CMASS galaxy samples across the redshift range $z = 0.34-0.7$ \citep{Anderson+2012,Ross+2014,Anderson+2014,Anderson+2014b,Burden+2014,Tojeiro+2014,Cuesta+2016,Gil-Marin+2016} and with the WiggleZ galaxies up to redshift $z\simeq 1$ \citep{Kazin+2014,Beutler+2016}, resulting in a substantial improvement on the measurement of the BAO scale.

The main limitation of this technique is the approximate treatment of the non-linear dynamics that may degrade the quality of the reconstruction. More sophisticated techniques are required in next-generation redshift surveys, such as those realised by PFS-SuMIRe \citep{PFSsumire}, eBOSS \citep{eBOSS}, 4MOST \citep{4MOST}, DESI \citep{DESI}, Euclid \citep{Euclid}, or WFIRST \citep{WFIRST}, which will sample a large range of galaxy overdensities within cosmological volumes. Indeed, in large overdensity regions (i.e. small scales) the linear approximation for the RSD correction adopted so far fails, the Zel'dovich approximation not being an exact solution \citep{Nusser+1991,Burden+2015}. Moreover, although optimal in recovering the BAO signature in the monopole of the correlation function or in the power spectrum also from large realistic samples \citep{KeselmanNusser2017}, the efficiency of the Lagrangian perturbative schemes to recover the BAO feature from higher-order multipoles and from the full 2D correlation functions, and simultaneously to account for the small scales for biased tracers, is less evident \citep{White2014}.

Several reconstruction techniques that adopt different approaches alternative to the standard one have been proposed. One group is represented by methods that still rely on (higher-order) Lagrangian perturbation theory. They implement iterative schemes, such as \textsc{piza} \citep{CroftGaztanaga1997} or \textsc{ztrace} \citep{MonacoEfstathiou1999}, include the gravitational tidal-field tensor and some linearisation scheme \citep{Kitaura+2012,KitauraAngulo2012}, apply a local transform to the density field \citep{Falck+2012,McCullagh+2013}, Wiener filtering to the large-scale forward displacement \citep{TassevZaldarriaga2012}, or an iterative FFT-method \citep{Burden+2015}. The second group is that of reconstruction methods based on Bayesian theory, which have been tested on galaxy mock catalogues \citep{KitauraEnsslin2008,JascheWandelt2013,Wang+2013}. Finally, there are the fully non-linear techniques, in which simplifying assumptions are not made on the dynamical state of the system but rather of the orbits of the objects. One example is based on the optimal Monge-Amp\`ere-Kantorovic (MAK) mass transportation problem \citep{Frisch+2002,Brenier+2003,Mohayaee+2003}. Successfully applied to reconstruct the peculiar velocity field of the Two Micron All-Sky Redshift Survey catalog \citep{Lavaux+2010}, the MAK technique is however computational expensive and therefore very likely limited to reconstructions within few hundreds Mpc \citep{Mohayaee+2006}.
Recently, an efficient non-linear reconstruction method has been proposed by \citet{KeselmanNusser2017}, based on a forward iterative scheme that uses standard $N$-body techniques for exactly solving the equation of motions.

In this paper we shall focus on the non-linear method based on the cosmological least action principle \citep[LAP;][]{peebles1989tracing}, which yield the full trajectory of tracers along with their velocities as a byproduct. Further developed by \citet{Peebles1994,Peebles1995} and \citet{ShayaPeeblesTully1995} to trace the dynamics of galaxies in the Local Universe, the LAP has been investigated against cold dark matter $N$-body simulations to estimate the collapsing history of haloes  by \citet{BranchiniCarlberg1994} and \citet{DunnLaflamme1995}. An efficient version dubbed Fast Action Minimization (FAM) method has been then implemented by \citet[hereafter \citetalias{nusser2000least}]{nusser2000least} to reconstruct the dynamics of galaxies the nearby Universe, finally accounting for redshift-space distortions \citep{BranchiniEldarNusser2002} and tidal field effects \citep{Romano-Diaz+2005}.
 
In this paper, we present an extended version of the original FAM algorithm, dubbed eFAM, coded in C++ language and designed for massive spectroscopic surveys such as Euclid, able to reconstruct the trajectories of $O(10^6)$ objects in generic background cosmologies and that we specifically apply to BAO reconstructions. Since the method provides multiple solutions for the orbits of the particles in virialised regions, the impact of non-linearities is minimised by reconstructing the trajectories of only collapsed haloes considered as point-like tracers, hence neglecting their internal structure altogether.

The plan of the paper is as follows. In the next section, we present the eFAM algorithm. Section~\ref{sec:BAOreconstruction} describes the analysis of large dark matter halo catalogues extracted from \textsc{deus-fur} $\Lambda$CDM simulations and the modelling of the RSD. The analysis of reconstructed haloes' orbits is performed and discussed in section~\ref{sec:fit}, focusing on the monopole moment of the two-point correlation function in real- and redshift-space and the anisotropic correlation function in redshift-space. The efficiency of the eFAM algorithm to recover the BAO scale and a comparison with a reconstruction based on the simple Zel'dovich approximation is also discussed. Section~\ref{sec:conclusions} is dedicated to the conclusions.

\vspace{-0.4cm}
\section{The Reconstruction Method}\label{sec:eFAM}

\subsection{The extended Fast Action Minimisation (eFAM) method}

The original FAM algorithm (\citetalias{nusser2000least}) has been developed to recover the past orbits of point-like particles both in real- and redshift-space in an Einstein-De Sitter universe. Here, FAM is extended to a generic cosmology defined by the Hubble parameter $H$ and the linear growth factor $D$ or the linear growth rate $f$, as functions of the scale factor $a$. The action of a set of $N$ collisionless equal-mass point-like particles with comoving coordinates $\{\mathbf{x}_i\}_{i=1,\dots,N}$ sampling in an unbiased way a volume $V$ and interacting only by gravity is, in the weak-field limit, 
\begin{eqnarray}
S&=&\sum_{i=1}^N\int_0^{D_\mathrm{obs}}\mathrm{d}D\left[fEDa^2\frac{1}{2}\left(\frac{\mathrm{d}\mathbf{x}_i}{\mathrm{d}D}\right)^2 \right. \\ \nonumber
&&\left.+\frac{3\Omega_{\mathrm{m}0}}{8\pi fEDa}\left(\frac{1}{\bar{n}_\mathrm{obs}a^3_\mathrm{obs}}\frac{1}{2}\sum_{j\neq i, j=1}^N\frac{1}{|\mathbf{x}_i-\mathbf{x}_j|}+\frac{2}{3}\pi\mathbf{x}_i^2\right)\right],
\label{eq:action}
\end{eqnarray}
where $D$ is used as time variable and $E=H/H_0$ the dimensionless Hubble parameter. Here the subscript ``0'' denotes quantities at present time and ``obs'' at redshift $z=z_\mathrm{obs}$. The mean number density of objects $\bar{n}_\mathrm{obs}=N/V$ determines the mean mass density in the volume, $\bar{\rho}$, and its mass density parameter  $\Omega_\mathrm{m0}$. The trajectories of objects are the solutions of the equations of motion deduced from a stationary action, $\delta S = 0$, subject to mixed boundary conditions as in FAM.

Differently from the original FAM algorithm that used a tree-code to calculate the gravitational force-field, eFAM makes use of \textsc{GyrfalcON} \citep{dehnen2002hierarchical}, a very efficient Poisson solver that optimally combines a tree-code and the fast multipole method (FMM). The FMM implements an improved multipole-acceptance criterion for the splitting or execution of the cell-cell interaction, and a symmetric calculation of the cell-cell interactions that conserves the total momentum.
The computational complexity is finally reduced to $O(N)$.
This is the major improvement of eFAM over FAM since it allows its application to larger datasets than its predecessor and, more specifically, capable of targeting the BAO reconstruction goal.

\subsection{Orbits parametrisation in generic cosmology}\label{subsec:paramrspace}

As in \citetalias{nusser2000least}, the trajectories $\{\mathbf{x}_i(D)\}_i$ are described by a linear combination of $M$ time-dependent basis functions $\{q_n(D)\}_n$ with unknown coefficients $\mathbf{C}_{i,n}$, viz.
\begin{equation}
\mathbf{x}_i(D)=\mathbf{x}_{i,\mathrm{obs}} +\sum_{n=0}^M\mathbf{C}_{i,n}q_n(D).
\end{equation}
The $q_n(D)$ are polynomials defined such that their derivatives $p_n(D)\equiv\mathrm{d}q_n(D)/\mathrm{d}D$ are the Jacobi polynomials satisfying the orthogonality condition
\begin{equation}\label{eq:orthogonalityBasis}
\int_0^{D_\mathrm{obs}}\mathrm{d}D\,w(D)p_np_m = \delta^K_{nm}A_n,
\end{equation}
with $\delta^K_{nm}$ the Kronecker delta and $A_n$ determined by the recurrence relations. Differently from \citetalias{nusser2000least}, the weight function $w(D)\equiv fEDa^2$ now to depends on the background cosmology and is modelled by $K(2D/D_\mathrm{obs}-2)^\alpha(2D/D_\mathrm{obs})^{\beta}$, with constant parameters $(K,\alpha,\beta)$ computed by an internal fitting procedure; see Appendix~\ref{appendix:Jacobi}.

For every term $n$, the mixed boundary conditions deduced from $\delta\mathbf{x}_i(D_\mathrm{obs})=0$ and $\lim_{D\to 0}\dot{D}\boldsymbol{\theta}_i(D) =0$ are
\begin{equation}\label{eq:boundarycond}
q_n(D_\mathrm{obs})=0\,, \quad 
\lim_{D\to 0}afHDp_n(D)=0\,,
\end{equation}  
in which the dependence on $D$ is omitted for clarity for all but the basis functions and their derivatives.
Denoting $\boldsymbol{\theta}_i\equiv\mathrm{d}\mathbf{x}_i/\mathrm{d}D=\mathbf{v}_i/fDH$ the rescaled peculiar velocity of the $i$-th particle and
\begin{equation}\label{eq:pec_acc}
\mathbf{g}_i \equiv -\frac{1}{\bar{n}_\mathrm{obs}a^3_\mathrm{obs}}\frac{1}{2}\sum_{j\neq i, j=1}^N\frac{\mathbf{x}_i-\mathbf{x}_j}{|\mathbf{x}_i-\mathbf{x}_j|^3}+\frac{4}{3} \mathbf{x}_i
\end{equation}
its peculiar acceleration, the stationary variations of the action with respect to $\mathbf{C}_{i,n}$ give
\begin{eqnarray}
0&=&\frac{\partial S}{\partial \mathbf{C}_{i,n}} = \int _0^{D_\mathrm{obs}}\mathrm{d}D\,w\boldsymbol{\theta}_i p_n + \int _0^{D_\mathrm{obs}}\mathrm{d}D \frac{3\Omega_{m,0}}{8\pi fEDa}\mathbf{g}_i q_n \nonumber \\
&=& \left[w\boldsymbol{\theta}_i q_n\right]_0^{D_\mathrm{obs}}-\int_0^{D_\mathrm{obs}}\mathrm{d}D\left[\frac{\mathrm{d}(w\boldsymbol{\theta}_i)}{\mathrm{d}D} -\frac{3\Omega_{\mathrm{m}0}}{8\pi fEDa}\mathbf{g}_i \right] q_n.
\end{eqnarray}
With the boundary conditions (\ref{eq:boundarycond}), these $N\times M$ equations correspond to the Euler-Lagrange equations obtained from $\delta S(\mathbf{x}_1,\dots,\mathbf{x}_N,\boldsymbol{\theta}_1,\dots,\boldsymbol{\theta}_N)=0$. This assures that the search for the stationary point of the action with respect to the coefficients $\mathbf{C}_{i,n}$ is equivalent to the one with respect to the whole trajectories.

\subsection{Redshift-space}
\label{subsec:paramsspace}

In redshift-space we introduce the comoving redshift coordinates of the $i$-th object as
\begin{equation}
\mathbf{s}_{i,\mathrm{obs}}=\frac{H_0a_0}{c}\mathbf{x}_{i,\mathrm{obs}}+\frac{a_0(fDH)_\mathrm{obs}}{c}\boldsymbol{\theta}^\parallel_{i,\mathrm{obs}},
\label{eq:redshift}
\end{equation}
in which $\boldsymbol{\theta}^\parallel_i$ denotes the component of the peculiar velocity along the line-of-sight. The additional term proportional to $\boldsymbol{\theta}^\parallel_{i,\mathrm{obs}}$, absent in real-space, breaks the isotropy of the $\mathbf{s}_{i,\mathrm{obs}}$ introducing a preferential direction along the line-of-sight; a Cartesian decomposition of the orbits coefficients $\mathbf{C}_{i,n}$ is therefore not convenient anymore. Instead, the coefficients $\mathbf{C}_{i,n}$ can be split into two components perpendicular and parallel to the line-of-sight. In the extended version of FAM we implemented the approach illustrated in \citet{SchmoldtSaha1998} by assigning to each object a Cartesian coordinate system with one axis aligned to the line-of-sight and the observer's position as the common origin of the galaxies' frames. In this way, the correction for the RSD is confined to one single axis, parallel to the radial velocity. Note that although the objects move, their coordinate frames do not. 

We adopted the orbits parametrisation introduced in \citetalias{nusser2000least} (see their equation~(20) and Appendix~\ref{appendix:s-space}) with
\begin{equation}\label{eq:boundarycond_delta_Q_s}
 Q_n(D_\mathrm{obs})=-(fDE)_\mathrm{obs}p_n(D_\mathrm{obs}),\quad n=1,\dots,M.
\end{equation}
To preserve the time-averaged equations of motion a kinetic energy term corresponding to a degree of freedom parallel to the line-of-sight is added to the action of the system; the resulting action in redshift-space to be minimised is
\begin{equation}\label{eq:action_sspace}
\mathcal{S}=S+\frac{1}{2}\left(w fDE\right)_\mathrm{obs}\left(\boldsymbol{\theta}_{i,\mathrm{obs}}^\parallel \right)^2.
\end{equation}

\subsection{Minimisation procedure and first guess}\label{subsec:firstguess}

The very hard minimisation problem in $3N\times M$ dimensions, which can be as large as $10^7$ for several millions of objects and $M\simeq10$ polynomials, is carried out using the same non-linear conjugate gradient method with the Polak-Ribi\`ere formula as in \citetalias{nusser2000least}.

However, this method is locally optimal whilst the action $\mathcal{S}$ can have many minima corresponding to different solutions of the time-averaged equations of motion. Indeed, as pointed out by \citet{peebles1989tracing} and \citet{giavalisco1993generalized} the solution of this mixed boundary-value problem is not unique, because the boundary conditions prescribe the time-dependence of the velocities near the initial time $D=0$ but do not specify their amplitude. Since we are here interested in the large-scale dynamics, the minimum should correspond to orbits that do not significantly deviate from the Zel'dovich approximation. The first guess of the iterative minimisation is then chosen as the one prescribed by the Zel'dovich approximation: the peculiar gravitational acceleration (\ref{eq:pec_acc}) linearly scales with the growth factor, $\mathbf{g}_i(D)=\frac{D/D_\mathrm{obs}}{a/a_\mathrm{obs}}\mathbf{g}_i(D_\mathrm{obs})$, and all but the zeroth-order coefficient are vanishing, i.e. $\mathbf{C}_{i,n}=0$ for all $n>0$, corresponding to straight-line orbits.

For both the real-space ($r$-space) and the perpendicular components in the redshift-space, the orthogonality condition (\ref{eq:orthogonalityBasis}) yields 
\begin{equation}
\mathbf{C}^{r-\mathrm{space}}_{i,n}=\mathbf{C}^\perp_{i,n}=-\frac{1}{A_n}\frac{a_\mathrm{obs}}{D_\mathrm{obs}}\mathbf{g}_i(D_\mathrm{obs})\int _0^{D_\mathrm{obs}}\mathrm{d}D \frac{3\Omega_\mathrm{m0}}{8\pi fEa^2} q_n,
\end{equation}
while for the parallel component in redshift-space
\begin{equation}
\mathbf{C}^\parallel_{i,n}=\frac{-a_\mathrm{obs}\mathbf{g}^\parallel_i(D_\mathrm{obs})}{[A_n+(w fDEp_n^2)_\mathrm{obs}]D_\mathrm{obs}}\int _0^{D_\mathrm{obs}}\mathrm{d}D \frac{3\Omega_\mathrm{m0}}{8\pi fEa^2} q_n.
\end{equation}

\vspace{-0.2cm}
\section{BAO reconstruction}\label{sec:BAOreconstruction}

\subsection{DEUS Full Universe Run haloes}\label{subsec:sim}

The eFAM method is formulated to reconstruct the complete trajectory of objects potentially
well into the non-linear regime. Accordingly, its accuracy needs then to be tested using simulated catalogues produced by fully nonlinear $N$-body experiments rather than those obtained with approximated schemes like \textsc{pthalos} \citep{PTHalos,PTHalos2}, \textsc{cola} \citep{COLA,L-PICOLA}, \textsc{pinocchio} \citep{PINOCCHIO,Munari+2017}, \textsc{patchy} \citep{PATCHY}, \textsc{ezmocks} \citep{EZmocks}, or \textsc{halogen} \citep{HALOGEN}, which are less accurate on small scales. Moreover, all these algorithms add random peculiar velocities to galaxies, typically drawn from a Maxwellian distribution scaled on the underlying matter density; this stochastic recipe is not compatible with the deterministic nature of the eFAM algorithm.

Aiming at assessing the quality of the reconstruction on large scales in the presence of significant non-linear effects, we have considered various sets of dark-matter haloes at redshift $z=0$, extracted from the Dark Energy Universe Simulation - Full Universe Run \citep[\textsc{deus-fur};][]{Rasera+2014}, a flat $\Lambda$CDM simulation set on the WMAP-7 best-fit cosmology \citep{Spergel+2007_WMAP7}\footnote{Cosmological parameters: $\Omega_\mathrm{b}=0.04356$, $\Omega_\mathrm{m}=0.2573$, $\Omega_\mathrm{r}=0.000049$, $h=0.72$, $n_\mathrm{s}=0.9630$, $\sigma_8=0.8010$.} employing $8192^3$ dark matter particles with formal mass and spatial resolution respectively of $1.2\times10^{12}h^{-1}M_\odot$ and $40h^{-1}$kpc, in a cubic volume of $(21 h^{-1}\mathrm{Gpc})^3$. The friend-of-friend (FoF) halo catalogue includes only haloes with more than 100 particles, amounting to more than 144 millions haloes at $z=0$. We stress that this is a demanding but somewhat unrealistic test, since next generation surveys will probe higher redshifts where nonlinear effects are less prominent.

The large volume of \textsc{deus-fur}, which encompasses the Hubble horizon thus enabling cosmic-variance limited predictions at the BAO scale, allows us to extract 512 cubic sub-volumes of length $2 h^{-1}$Gpc separated by a buffer region of $500 h^{-1}$Mpc, a distance above the scale of homogeneity (see e.g. \citealt{Ntelis+2017} for the results on the BOSS DR12 galaxy sample). The sub-volumes can therefore be considered as effectively independent, allowing for a Monte Carlo estimation of the covariance \citep{Norberg+2009}.

The peculiar velocities of haloes are not supplied with the \textsc{deus-fur} FoF catalogue. To emulate the RSD and perform the reconstruction in redshift-space, eFAM is run a first time on every sub-volume to assign the peculiar velocities to haloes.
Then for every sub-volume a reference frame is fixed in its centre-of-mass, the redshift coordinates of haloes are computed using equation (\ref{eq:redshift}), and the action (\ref{eq:action_sspace}) is minimised. 
The accuracy of this procedure is discussed and quantitatively assessed in Section~\ref{subsec:vvcomparison}

\vspace{-0.2cm}
\subsection{Gravitational tidal-field}\label{subsec:tidal}

In its basic formulation the FAM method does not assume any external gravitational field, treating the sampled density field as an isolated system. This assumption does not apply in a cosmological context, however the impact of the external density field can be minimised by choosing a spherical geometry for the sample. Here the reconstruction is applied to spherical domains $\mathcal{D}$ of radius $990h^{-1}$Mpc extracted from each sub-volume, each containing about 56,000 haloes on average.

A direct computation of the external gravitational field $\Phi_\mathrm{tidal}(\mathbf{x})$ affecting the dynamics inside the sub-region $\mathcal{D}_\mathrm{int}$ can be performed by extending the reconstruction to a larger domain $\mathcal{D}$ embedding $\mathcal{D}_\mathrm{int}$, as done by \cite{ShayaPeeblesTully1995}. To assess the impact of $\Phi_\mathrm{tidal}(\mathbf{x})$, we applied eFAM to haloes in several concentric spherical domains with radius ranging from $100h^{-1}$Mpc to $300h^{-1}$Mpc, extracted from a smaller \textsc{deus} simulation for which the peculiar velocities are provided. We then performed a point-wise comparison between the Cartesian components of reconstructed and real velocities of the haloes that are in the common domain $\mathcal{D}_\mathrm{int}$ of radius $100h^{-1}$Mpc, and assessed the impact of $\Phi_\mathrm{tidal}(\mathbf{x})$ by the offset $q$ of the linear regression $v_\mathrm{FoF}=mv_\mathrm{eFAM}+q$. It turns out that the reconstruction within $\mathcal{D}_\mathrm{int}$ is improved by the inclusion of a buffer region consisting of a spherical shell of at least $200h^{-1}$Mpc, reducing the offset from $q\sim 100$km\,s$^{-1}$ when no external field is considered to $q\sim 10$km\,s$^{-1}$, regardless of the Cartesian component. Therefore, in the \textsc{deus-fur}-based BAO reconstruction, we shall only analyse the regions within $700h^{-1}$Mpc, each containing about 23,000 haloes on average, and ignore the outskirts that extend to $990h^{-1}$Mpc.

\vspace{-0.5cm}
\section{Fitting the acoustic feature}\label{sec:fit}

\subsection{Measuring, modelling, and fitting the correlation function}\label{subsec:mesmodfit}

For each sub-sphere $\mathcal{D}_\mathrm{int}$ the monopole of the two-point correlation function $\xi(r)$ is computed in the separation range $30-200h^{-1}$Mpc with linear binning of 10$h^{-1}$Mpc using the minimum variance \citet{LandySzalay1993} estimator, with 50 times random objects homogeneously distributed within $\mathcal{D}_\mathrm{int}$. The measurement is repeated for all sources at their pre- and post-reconstructed positions at 14 different redshifts, starting from $z=0$ and up to the maximum redshift $z_\mathrm{max}$ allowed by reconstruction, defined as the redshift for which $\Sigma_\mathrm{NL}$ attains the minimum positive value.

The results are fitted using the \citet{XuPadmanabhan+2012} model
\begin{equation}
\xi(r;z)=B^2\xi_\mathrm{m}(\alpha r;z)+A(r),
\end{equation}
where $B$ is a multiplicative constant bias and $\xi_\mathrm{m}(r)$ is the Fourier transform of
\begin{equation}
P(k)=[P_\mathrm{lin}(k)-P_\mathrm{smooth}(k)]\,\mathrm{e}^{-k^2\Sigma_\mathrm{NL}^2/2}+P_\mathrm{smooth}(k),
\end{equation}
with $P_\mathrm{lin}(k)$ the actual linear power spectrum and $P_\mathrm{smooth}(k) $ its de-wiggled limit, both computed by \textsc{camb} \citep{CAMB} using the same cosmological model adopted in the \textsc{deus-fur} simulations. The broad-band term
\begin{equation}
A(r)=A_0+\frac{A_1}{r}+\frac{A_2}{r^2}
\end{equation}
can be interpreted as an effective description of mode-coupling \citep[e.g.][]{CrocceScoccimarro2008} not affecting the BAO scale but biasing its measurement if not accounted for properly. This additive term can also help to alleviate the effects of assuming a wrong cosmological model. The broadening and shift of the BAO feature due to the non-linear growth of structures are described by the parameter $\Sigma_\mathrm{NL}$, accounting for the Lagrangian displacements, and by the scale dilation parameter $\alpha$. The model has therefore six free parameters, $(B^2,\alpha,\Sigma_\mathrm{NL},A_0,A_1,A_2)$, and fixed cosmological parameters.

The data are fitted using a MCMC technique with Gaussian likelihood and flat priors. The covariance matrix is calculated from the $N_\mathrm{mocks}=512$ mocks as
\begin{equation}\label{eq:cov}
\mathbfss{C}_{i,j}=\frac{1}{N_\mathrm{mocks}-1}\sum_{n=1}^{N_\mathrm{mocks}}[\xi_n(r_i)-\bar{\xi}(r_i)][\xi_n(r_j)-\bar{\xi}(r_j)]
\end{equation}
where $\bar{\xi}(r)=\sum_{n=1}^{N_\mathrm{mocks}}\xi (r)/(N_\mathrm{mocks}-1)$ is the mean two-point correlation function.

Measurements, modelling, and fitting are performed using the routines implemented in the \textsc{CosmoBolognaLib} \citep{Marulli+2016cosmobolognalib}.

\subsection{Monopole in real space}\label{subsec:monopole}

\begin{figure}
 \includegraphics[width=0.9\columnwidth]{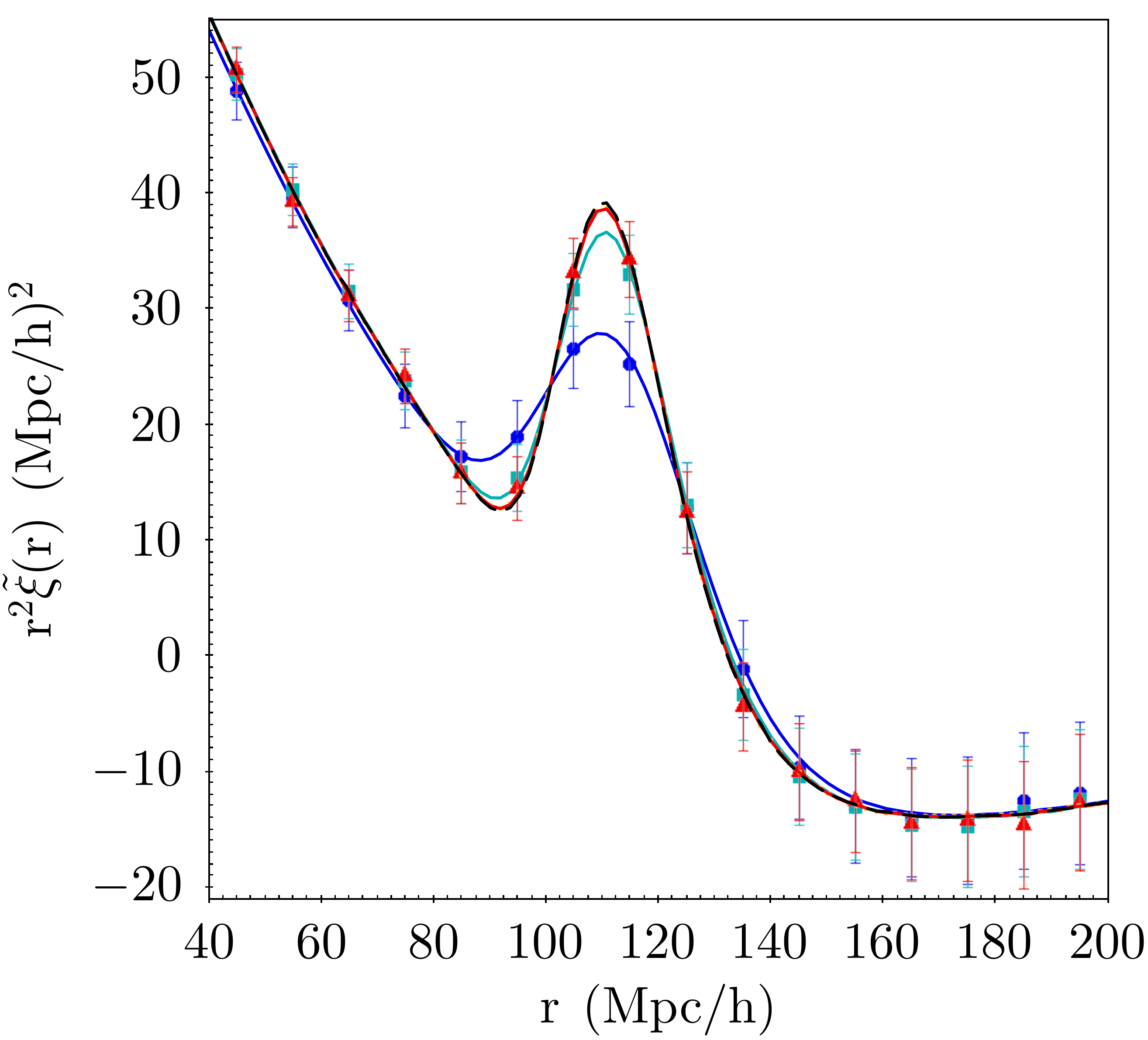}
\caption{Rescaled two-point correlation function in real-space of \textsc{deus-fur} $\Lambda$CDM dark-matter haloes before and after reconstruction by eFAM ($M=10$), averaged over the 512 mocks. Before reconstruction at redshift $z=0$ (blue line, circles) the acoustic feature is broadened by non-linear evolution and peculiar velocities. After reconstruction at $z=2.7$ and $z=6.5$ (green/squares and the red/triangles, respectively) the measured correlation function gets progressively closer to the the linear prediction (black-dashed), indicating the quality of the eFAM algorithm. Error bars from the diagonal of the covariance matrix.}
\label{fig:eFAM_xi_med_rspace}
\end{figure}

\begin{figure*}
\centering
\subfloat
   {\includegraphics[width=0.7\columnwidth]{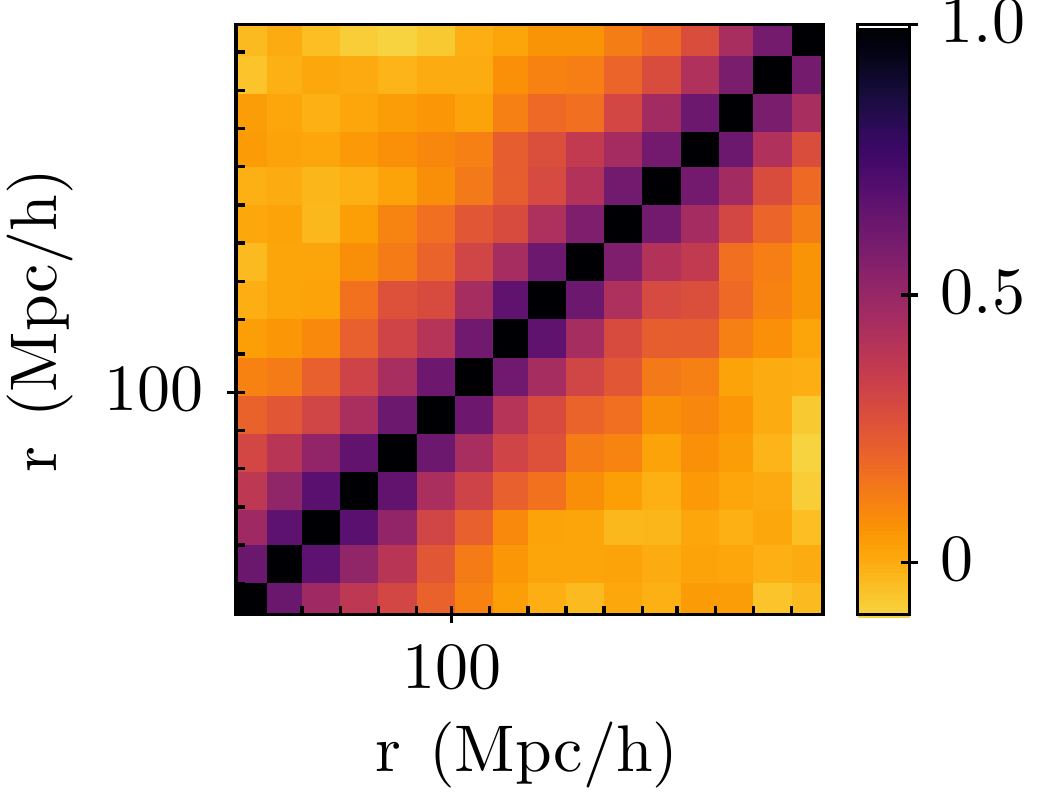} } 
\subfloat
   {\includegraphics[width=0.7\columnwidth]{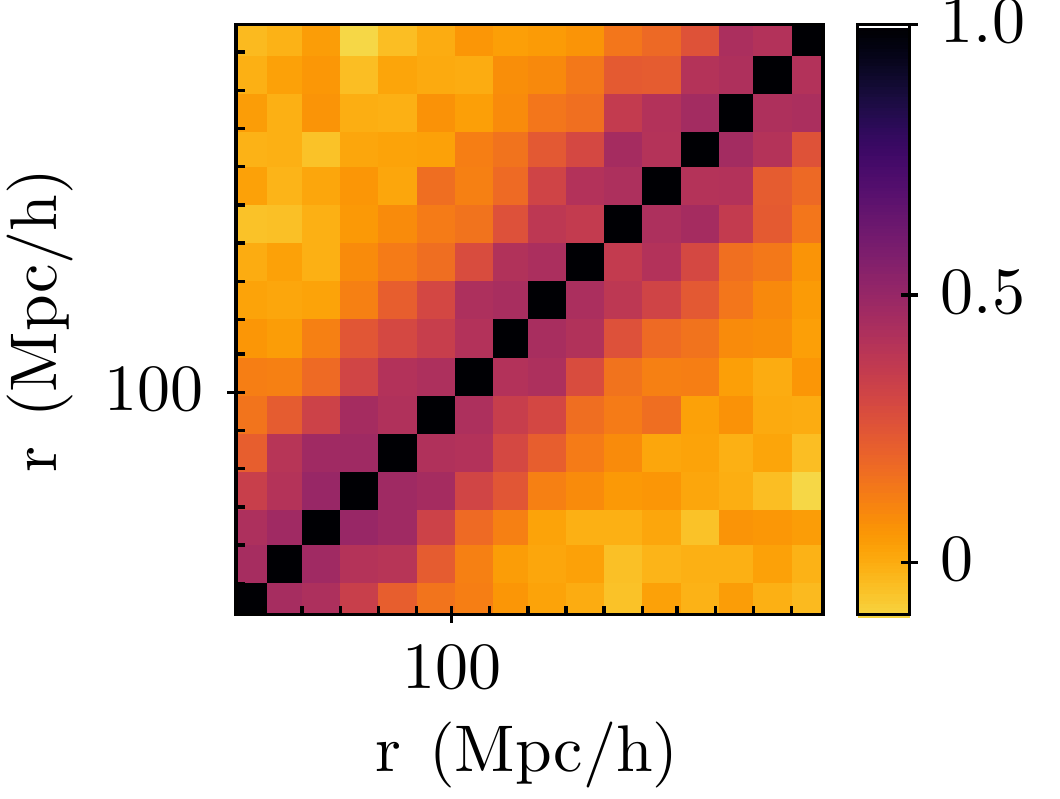}} 
\subfloat
   {\includegraphics[width=0.7\columnwidth]{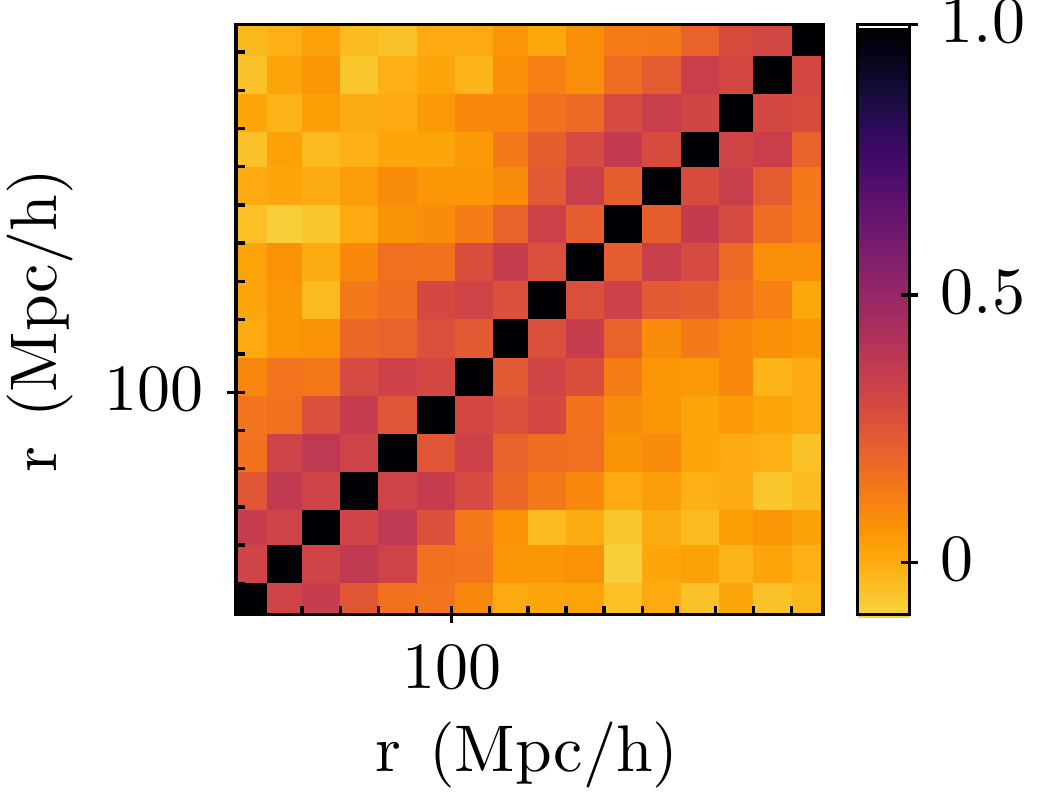}}
\caption{Correlation matrix of the two-point correlation function (monopole, real space) around the BAO scale. Left: from non reconstructed halo catalogues. Middle and right: from reconstructed haloes, using eFAM$_{10}$, at $z=2.7$ and $z=6.5$, respectively. The correlation matrix becomes more diagonal at higher redshifts, proving that the correlation functions of the different mocks tends to converge to the mean form when reconstructed.}
\label{fig:Corr_matrix_rspace}
\end{figure*}

An ideal reconstruction pushed at early time, before the non-linear clustering became effective, would yield $\Sigma_\mathrm{NL}\to0$ and $\alpha\to1$. As shown in Figure~\ref{fig:eFAM_xi_med_rspace}, using the eFAM algorithm with $M=10$ basis functions (hereafter quoted as eFAM$_{10}$) the template model $\xi_\mathrm{m}(r)$ based on the linear theory is closely approached at redshift $z=2.7$ and almost fully restored at $z=6.5$ (to ease the comparison between the correlation functions at different redshifts and enhance the acoustic feature, the rescaled monopole $r^2\tilde{\xi}(r;z)=r^2\left[\xi(r;z)-A(r))\right][B^2D(z)^2]^{-1}$ is plotted). The error bars are the rms-variance obtained from the diagonal elements of the covariance matrix (\ref{eq:cov}). Remarkably enough, the errors on the monopole calculated pre-reconstruction do not increase after reconstruction. More interestingly, the correlation matrix becomes definitely more diagonal going towards higher redshift; see Figure~\ref{fig:Corr_matrix_rspace}. This indicates that the reconstruction de-correlates the signal in all bins.

The fact that the linear regime is almost fully restored already at $z=6.5$ rather than higher redshifts, as expected, reflects the fact that eFAM overestimates the amplitude of peculiar velocities of objects in high-density environments (small scales), where linear theory fails, but predicts their correct direction. As a result, the back in time displacement of these objects is overestimated, the density contrast is more efficiently reduced and, consequently, linear theory restored at lower redshifts than expected. Such overcorrection does not affect the quality of the BAO reconstruction as long as it does not lead to an un-physical compression of the BAO peak, namely below the Silk scale.

\begin{figure*}
\centering
\subfloat
   {\includegraphics[width=0.9\columnwidth]{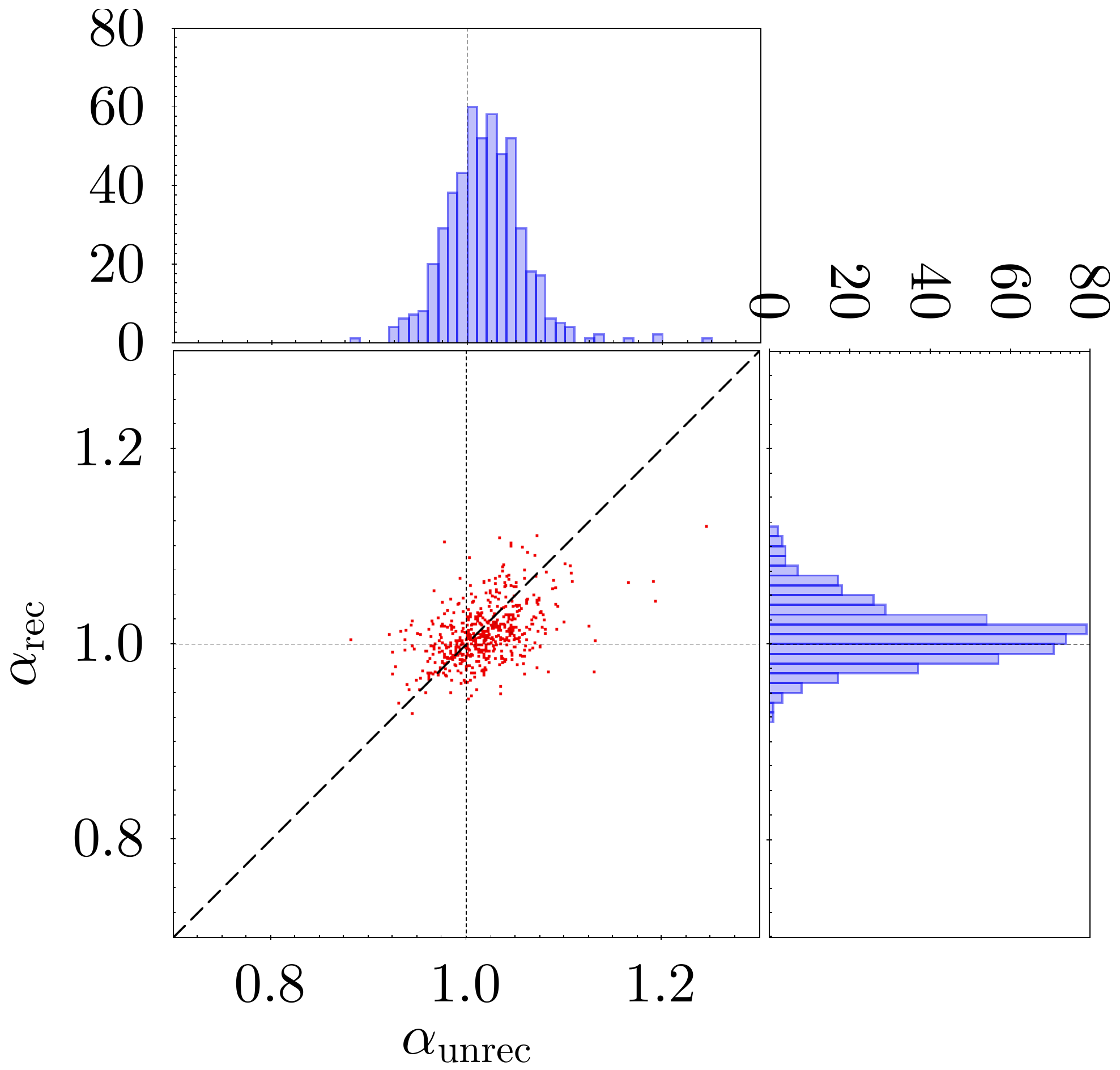}
   \put(-179,192){$\langle\alpha_\mathrm{unrec}\rangle = 1.01 $}
   \put(-179,182){$\sigma_{\alpha,\mathrm{unrec}} = 0.04$ }
   \put(-62,129){$\langle\alpha_\mathrm{rec}\rangle = 1.00 $}
   \put(-62,119){$\sigma_{\alpha,\mathrm{rec}} = 0.03$ }} \quad \quad \quad \quad
\subfloat
   {\includegraphics[width=0.9\columnwidth]{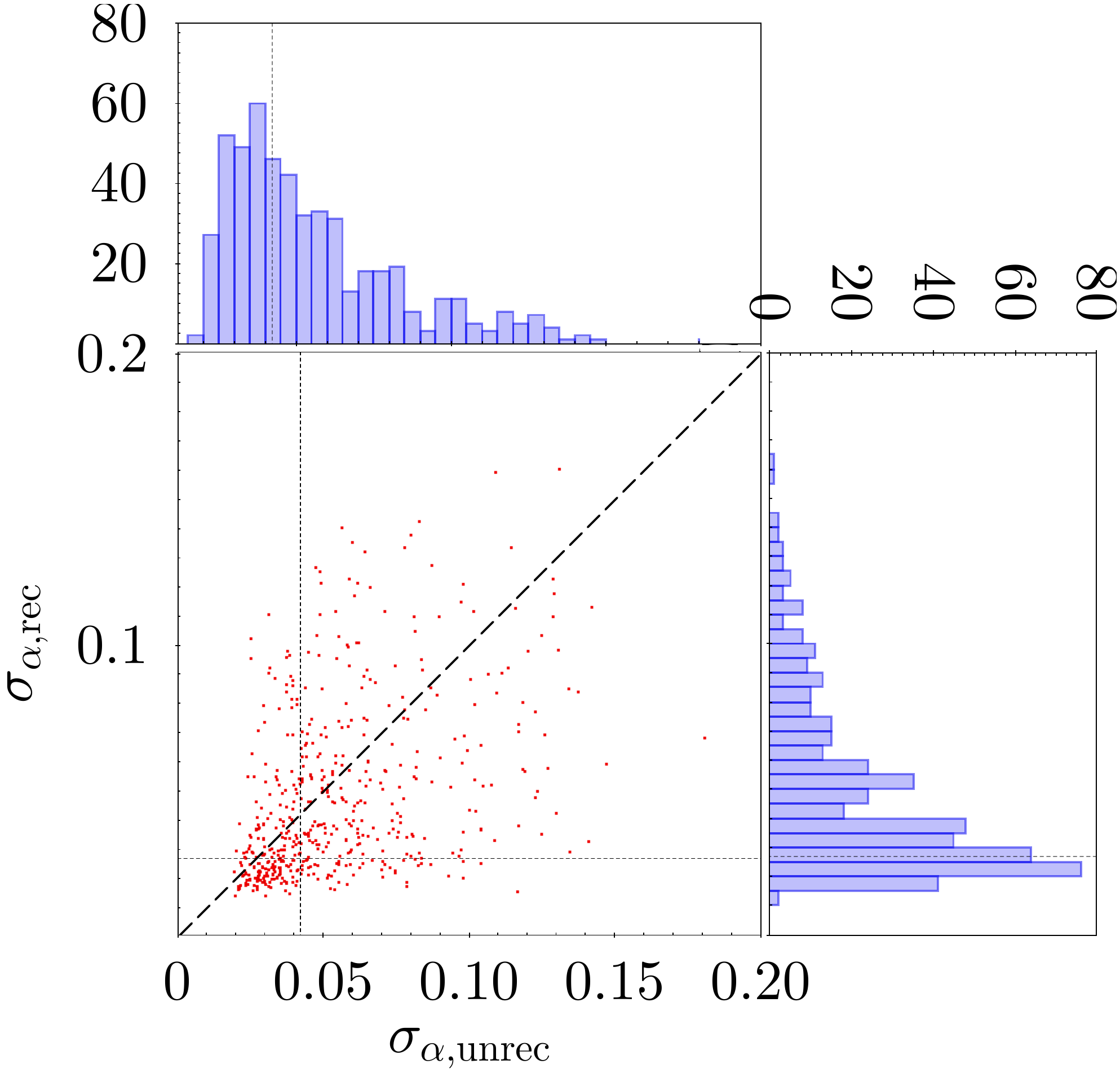}
   \put(-150,120){31\% of points}
   \put(-125,50){69\% of points}} 
\caption{Comparison of scale dilation parameter $\alpha$ (left panel) and its error (right panel) estimated from the real-space mocks pre- (``unrec'') and post-reconstruction (`rec''), together with the marginalised histograms, for the 512 mocks. {\it Left:} the black-dotted lines show the true values ($\alpha= 1$), the diagonal black-dashed line marks the perfect reconstruction ($\alpha_\mathrm{rec} = \alpha_\mathrm{unrec}$); the eFAM reconstruction reduces the scatter in these values without introducing any statistical bias. {\it Right:} short dashed line marks the equality of errors on the scale dilation parameter pre- and post-reconstruction, $\sigma_{\alpha,\mathrm{rec}} = \sigma_{\alpha,\mathrm{unrec}}$, the black-dotted lines correspond to the median values for the marginalised distributions; for 69 percent of mocks the error $\sigma_\alpha$ decreases after reconstruction.}
\label{fig:histo_rspace}
\end{figure*}
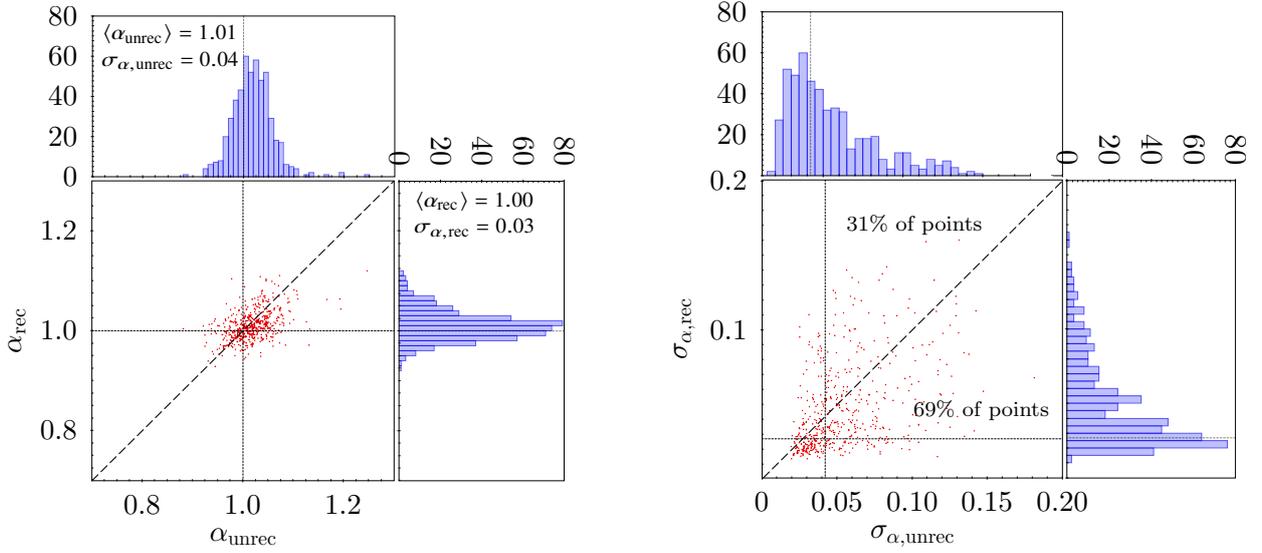

To estimate the impact of the reconstruction on the dilation $\alpha$ and its error, we performed the same analysis as in \citet{Padmanabhan+2012}: the two-point correlation function is fitted for every mock, using flat priors for all the parameters but $\Sigma_\mathrm{NL}$, for which a Gaussian prior centred on the best-fit obtained for the average $\bar{\xi}(r)$ and with the same variance. The scatterplot in the left panel of Figure~\ref{fig:histo_rspace} compares the values of $\alpha$ pre- and post-reconstruction for all the 512 halo catalogues. The non-linear eFAM method improves the measurement of the BAO scale reducing the standard deviation of the probability distribution function of $\alpha$, without introducing any statistical bias. The improvement of the precision on $\alpha$ (right panel) is significant, the eFAM algorithm yielding $\sigma_{\alpha,\mathrm{rec}}<\sigma_{\alpha,\mathrm{unrec}}$ for the 69 percent of mocks.

\begin{figure}
 \includegraphics[width=0.87\columnwidth]{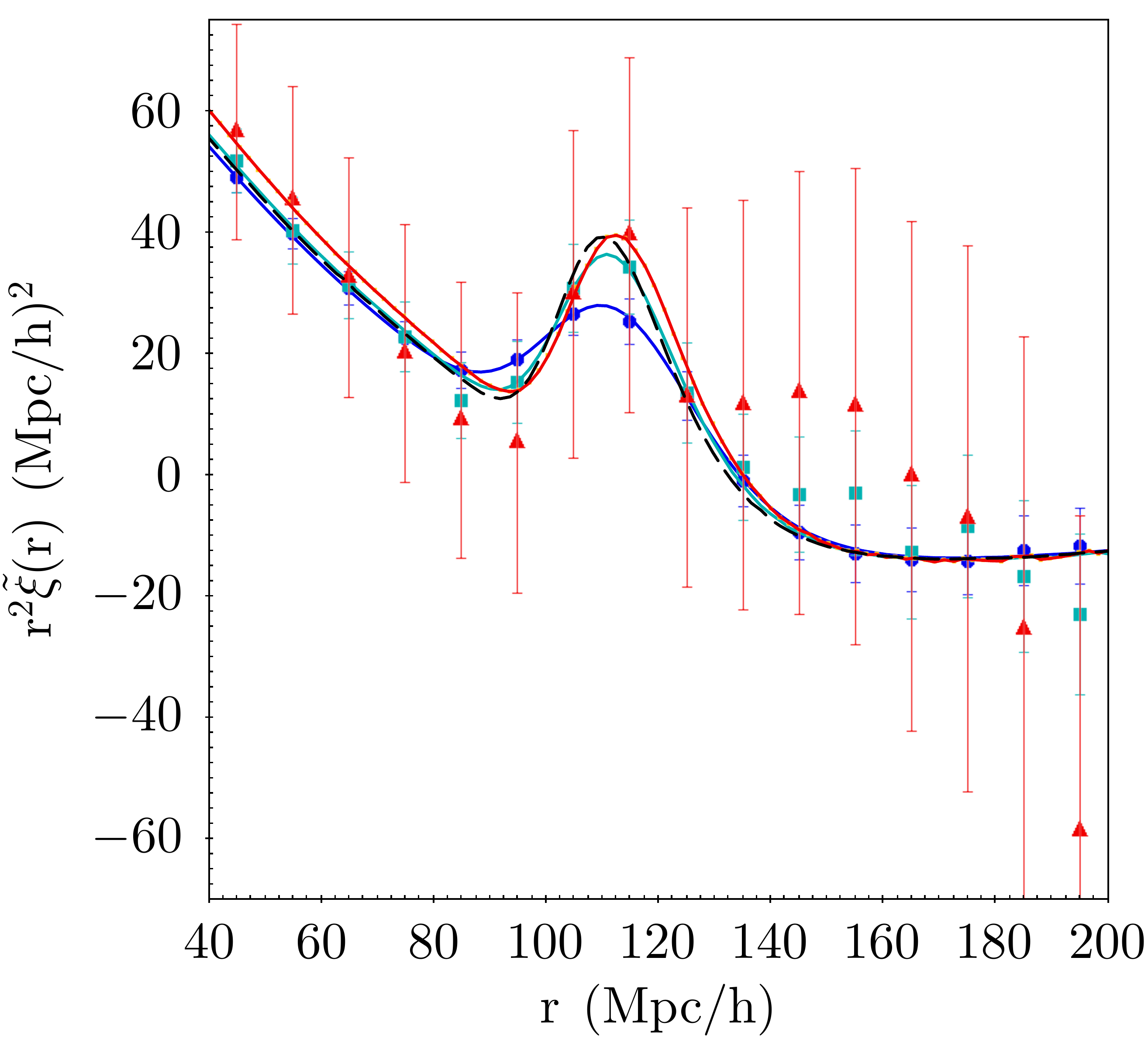}
\caption{Similar to Figure~\ref{fig:eFAM_xi_med_rspace} (real-space), but with reconstruction by the Zel'dovich approximation as provided by the eFAM with $M=1$. The very large errors indicate the non-reliability of this method, which moreover returns a biased estimation of the acoustic scale toward larger values.}
\label{fig:ZA_xi_med_rspace}
\end{figure}

The eFAM reconstruction is superior to the standard Zel'dovich one. To quantify the improvement we repeated the reconstructions using eFAM with $M=1$ (i.e. eFAM$_1$), that is forcing straight orbits. This is not strictly Zel'dovich approximation since velocities change along the orbit, but is a good approximation to it. As shown in Figure~\ref{fig:ZA_xi_med_rspace}, the averaged correlation function has a very large variance, with the acoustic peak systematically shifted toward larger values, $\alpha=0.98\pm0.01$. The poor reliability of the results achieved by eFAM$_1$ is mainly due to the low value of its maximum allowed redshift, $z=3.7$, above which the best-fit value of $\Sigma_\mathrm{NL}$ becomes unphysical; see table~\ref{tab:mean_xi_rspace}. This also explains why the results obtained at $z=2.7$ are better than those at $z=6.5$, which is the maximum allowed redshift for eFAM$_{10}$. Instead, owing to the larger number of degrees-of-freedom, at the same redshift the fully non-linear eFAM$_{10}$ method ensures a non-biased measurement of the acoustic scale, $\alpha=1.000\pm0.001$, and a non-linear broadening reduced to $\Sigma_\mathrm{NL}=1.2\pm0.7h^{-1}$Mpc. This value is smaller by a factor $\sim1.7$ than the one obtained with eFAM$_1$, which is moreover totally dominated by errors.

\begin{table}
	\centering
	\caption{Fit results to the average correlation function before and after reconstruction in real-space by eFAM$_{10}$ and eFAM$_1$. The parameters not shown are marginalised over. A negative (unphysical) best-fit value of $\Sigma_\mathrm{NL}$ indicates that $z_\mathrm{max}$ has been attained, values for $z > z_\mathrm{max}$ (marked by $^*$) are shown just for comparison.}
	\begin{tabular}{lccc}
		\hline
		Method & Redshift & $\alpha$
		 & $\Sigma_\mathrm{NL} \; (h^{-1}\mathrm{Mpc})$ \\
		\hline\hline
		pre-recon	& $z=0$ & 1.007 $\pm$ 0.002 & 9.0 $\pm$ 0.2 \\
		\hline
		eFAM$_{10}$ & $z= 2.7$ & 0.999 $\pm$ 0.001 &  3.5 $\pm$ 0.4\\
		 			& $z=3.7$ & 0.998 $\pm$ 0.001 & 2.3  $\pm$ 0.7\\
		 			& $z=4.3$ & 0.999 $\pm$ 0.001 & 1.6  $\pm$ 0.7\\
		  			& $z=6.5$ & 1.00 $\pm$ 0.001 & 1.2  $\pm$ 0.7\\
		\hline
		eFAM$_{1}$ & $z= 2.7$ & 0.996 $\pm$ 0.003 & 3.8  $\pm$ 1.2 \\
				   & $z=3.7$ & 0.997$\pm$ 0.003 & 2.0 $\pm$ 1.4 \\
				   & $z=4.3^*$ & 0.927 $\pm$ 0.018 & -3.4  $\pm$ 4.3\\
				   & $z=6.5^*$ & 0.980 $\pm$ 0.010 & 2.0 $\pm$ 2.1 \\
		\hline
	\end{tabular}
\label{tab:mean_xi_rspace}
\end{table}

\subsection{Recovering the BAO signal in statistically anomalous samples}\label{subsec:anomalous}

The analysis of the two-point correlation function before reconstruction shows that, in some sample, the scale of the BAO is badly constrained. We identified two types of samples: those returning a wrong best-fit $\alpha_\mathrm{unrec}$ typically deviating from the actual value $\alpha=1$ more than $\sigma_\alpha$ (dubbed type-I samples), with a corresponding $\chi^2(\alpha)$ with a minimum significantly shifted from the true value; and samples without a clear acoustic feature (type-II samples), often yielding a $\chi^2(\alpha)$ with a very shallow minimum. The eFAM method is remarkably able to recover the correct $\alpha$ value from both types of anomalous samples.

\begin{figure*}
\subfloat
   {\includegraphics[width=0.7\columnwidth]{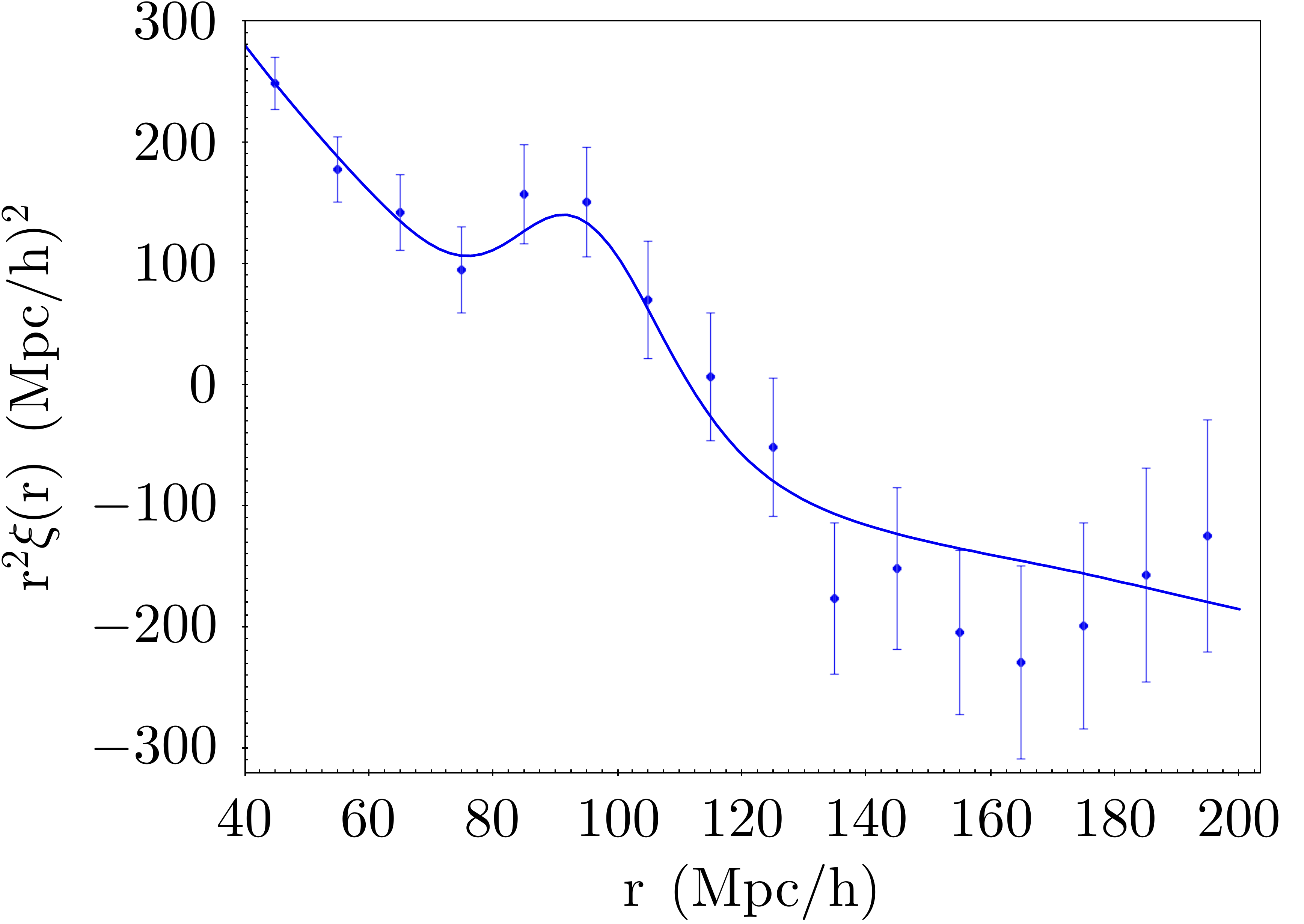}
    \put(-130,+40){$\alpha = 1.19$}
   \put(-130,30){$\sigma_\alpha = 0.07$ }} 
\subfloat
   {\includegraphics[width=0.7\columnwidth]{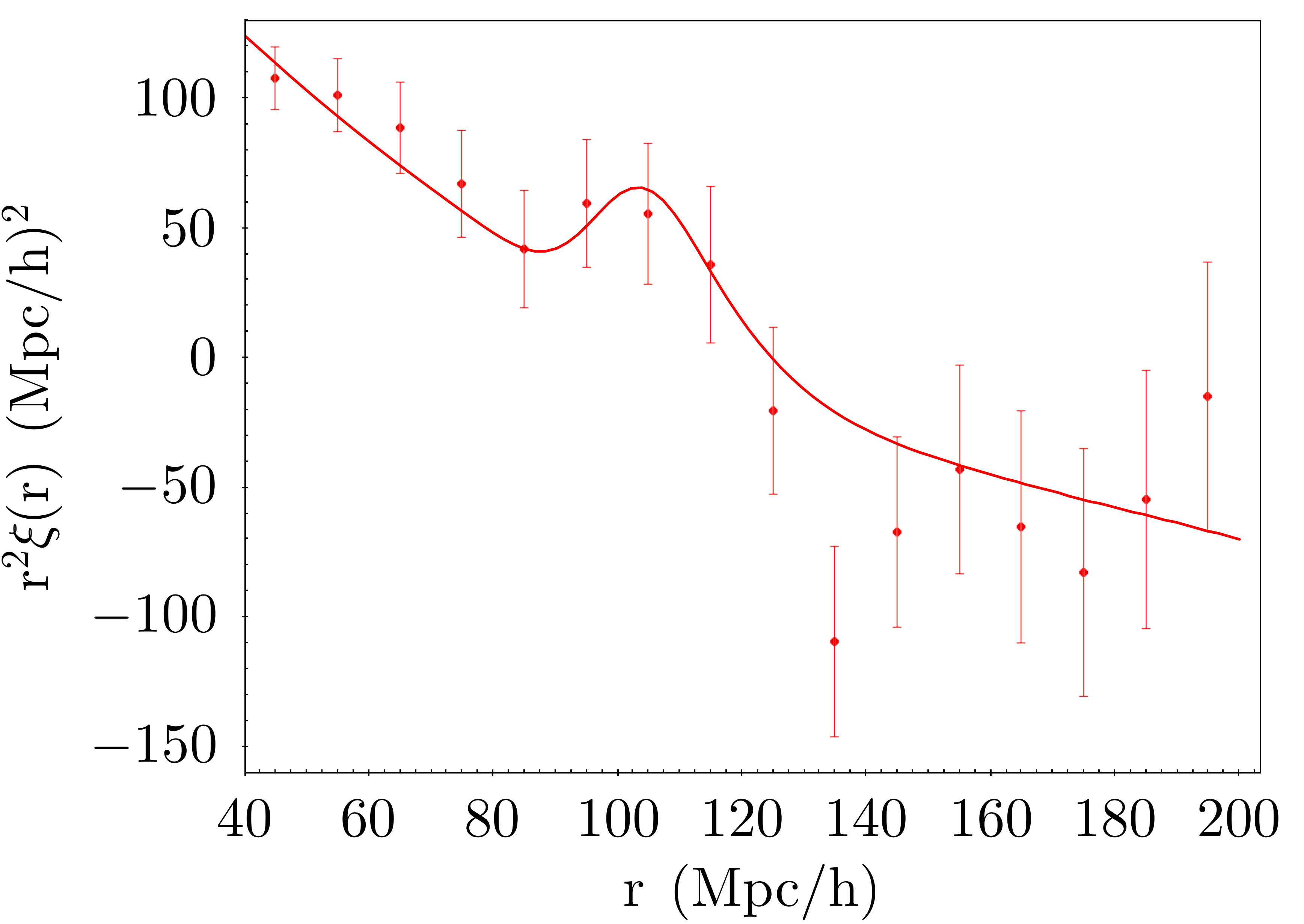}
    \put(-130,+40){$\alpha =1.06$}
   \put(-130,30){$\sigma_\alpha = 0.06$ }} 
\subfloat
   {\includegraphics[width=0.69\columnwidth]{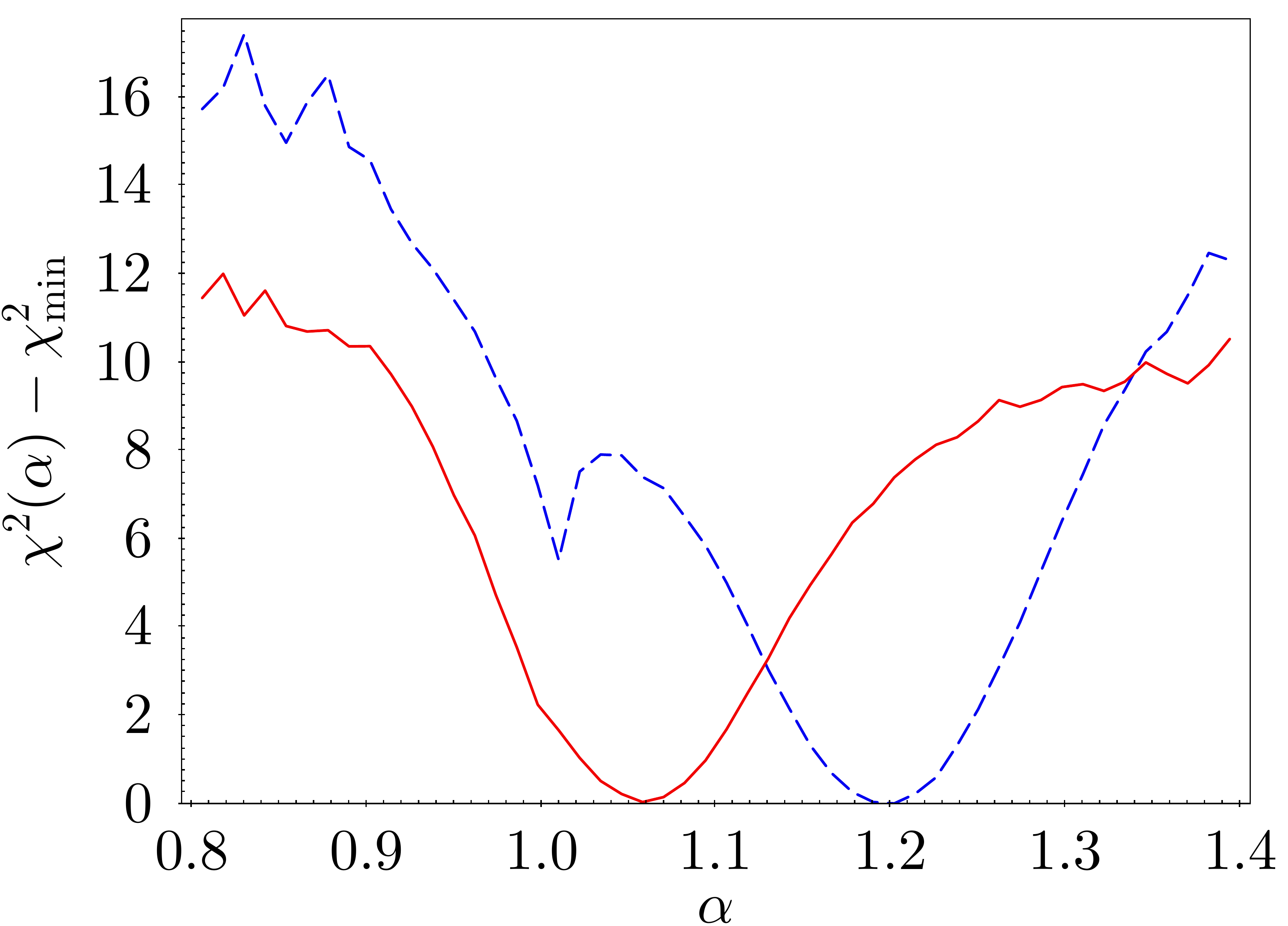}
   \put(-55,30){$\chi^2_\mathrm{min}=19.0$ }}\\
 \subfloat
   {\includegraphics[width=0.7\columnwidth]{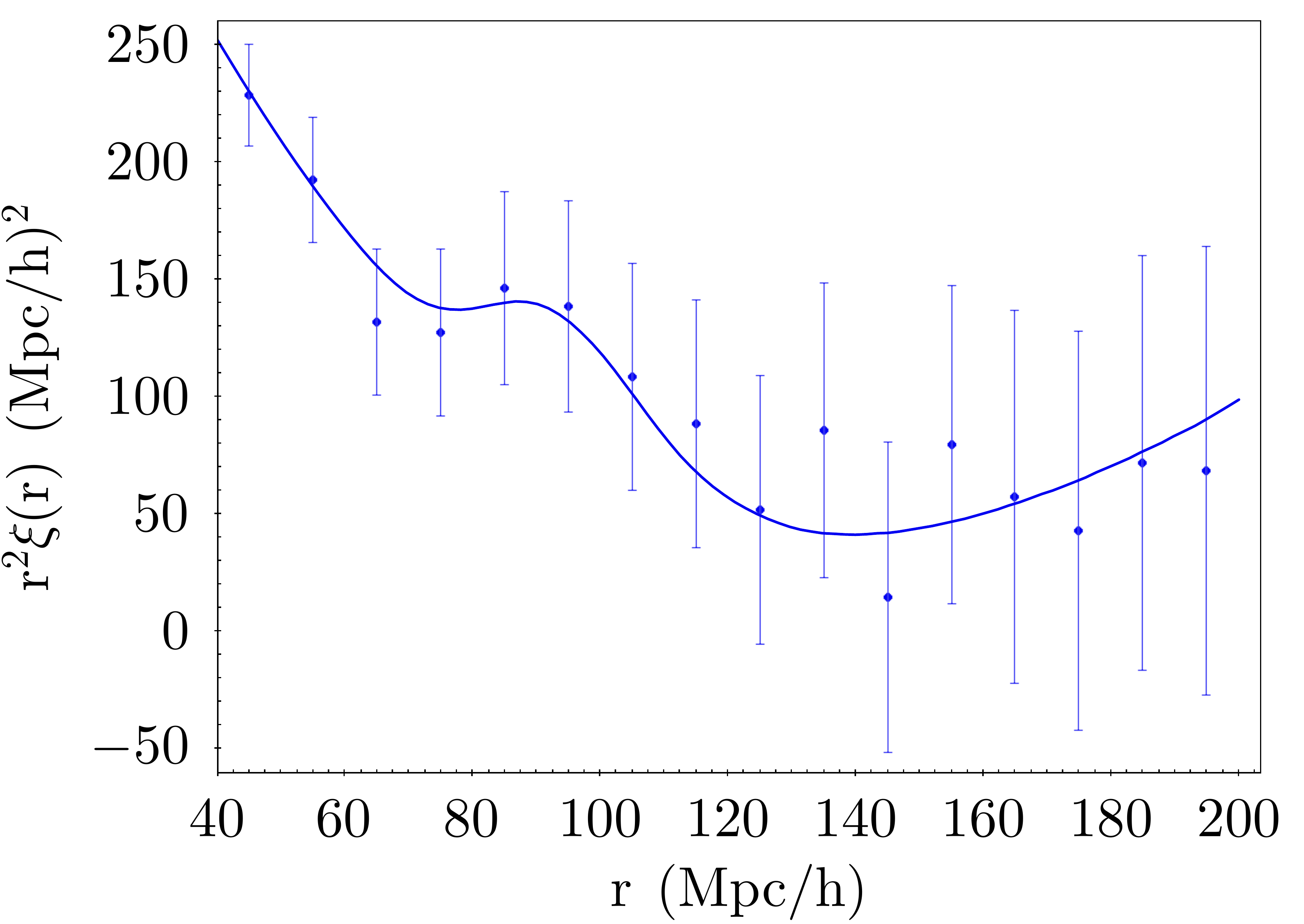}
    \put(-130,+40){$\alpha = 1.19$}
   \put(-130,30){$\sigma_\alpha = 0.11$ } } 
\subfloat
   {\includegraphics[width=0.7\columnwidth]{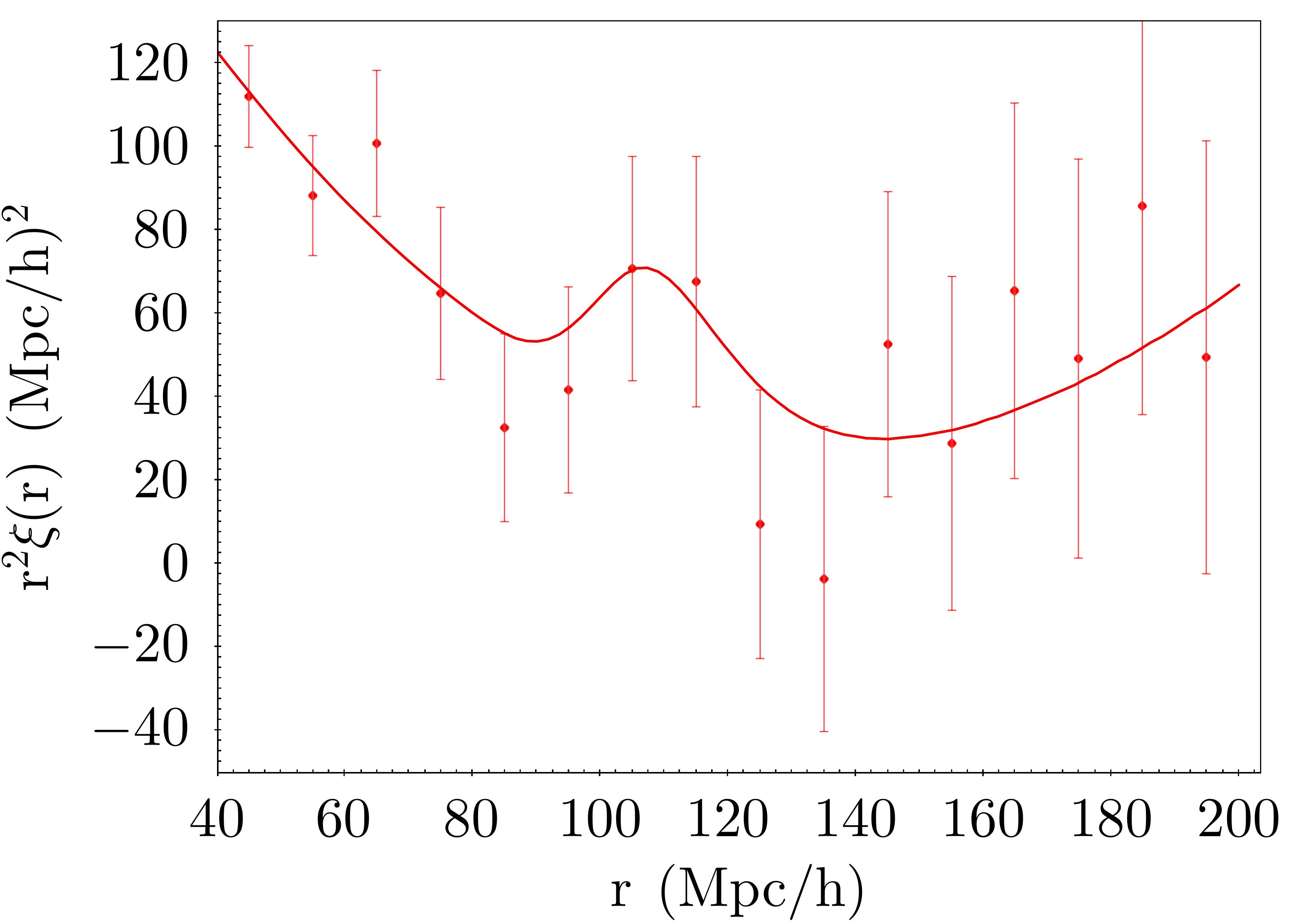}
    \put(-130,+40){$\alpha = 1.04$}
   \put(-130,30){$\sigma_\alpha = 0.08$ }} 
\subfloat
   {\includegraphics[width=0.69\columnwidth]{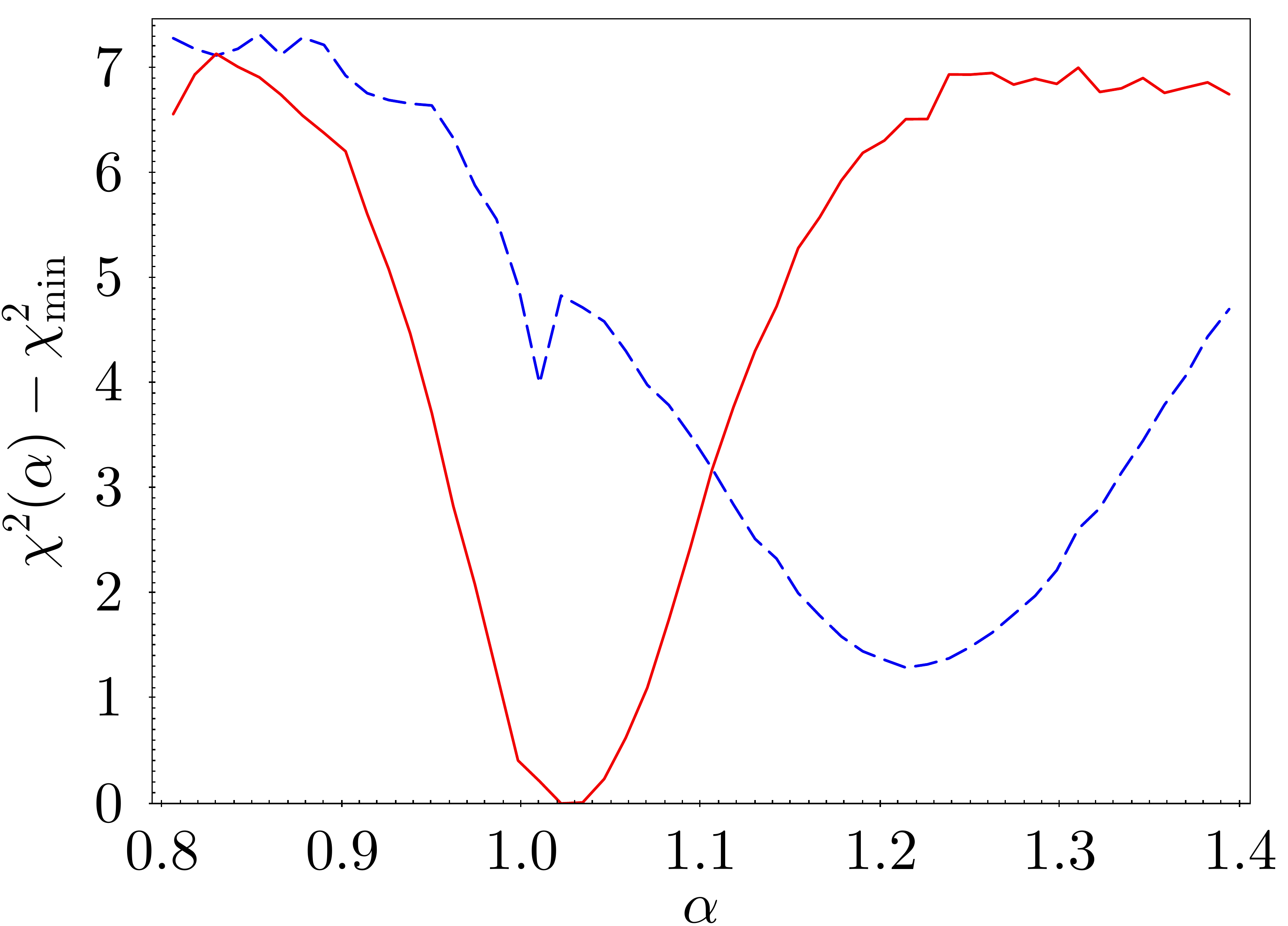}
   \put(-55,30){$\chi^2_\mathrm{min}=18.6$ }}\\
\caption{Fit results from the fiducial model for two type-I anomalous samples.
{\it Left column:} Monopole of the two-point correlation function before reconstruction; the acoustic peak is shifted towards small scales. {\it Middle column:} Correlation function after reconstruction; the peak location is now compatible the expected value within one standard deviation $\sigma_\alpha$. {\it Right column:} $\Delta\chi^2(\alpha)=\chi^2(\alpha)-\chi^2_\mathrm{min}$ before reconstruction (blue-dashed line) and after reconstruction (red line); the shift of the best-fit $\alpha$ towards the expected value $\alpha=1$ reflects the the shift of the peak in the correlation function.}
\label{fig:Unl_shift_rspace}
\end{figure*}

\begin{figure*}
 \subfloat
   {\includegraphics[width=0.7\columnwidth]{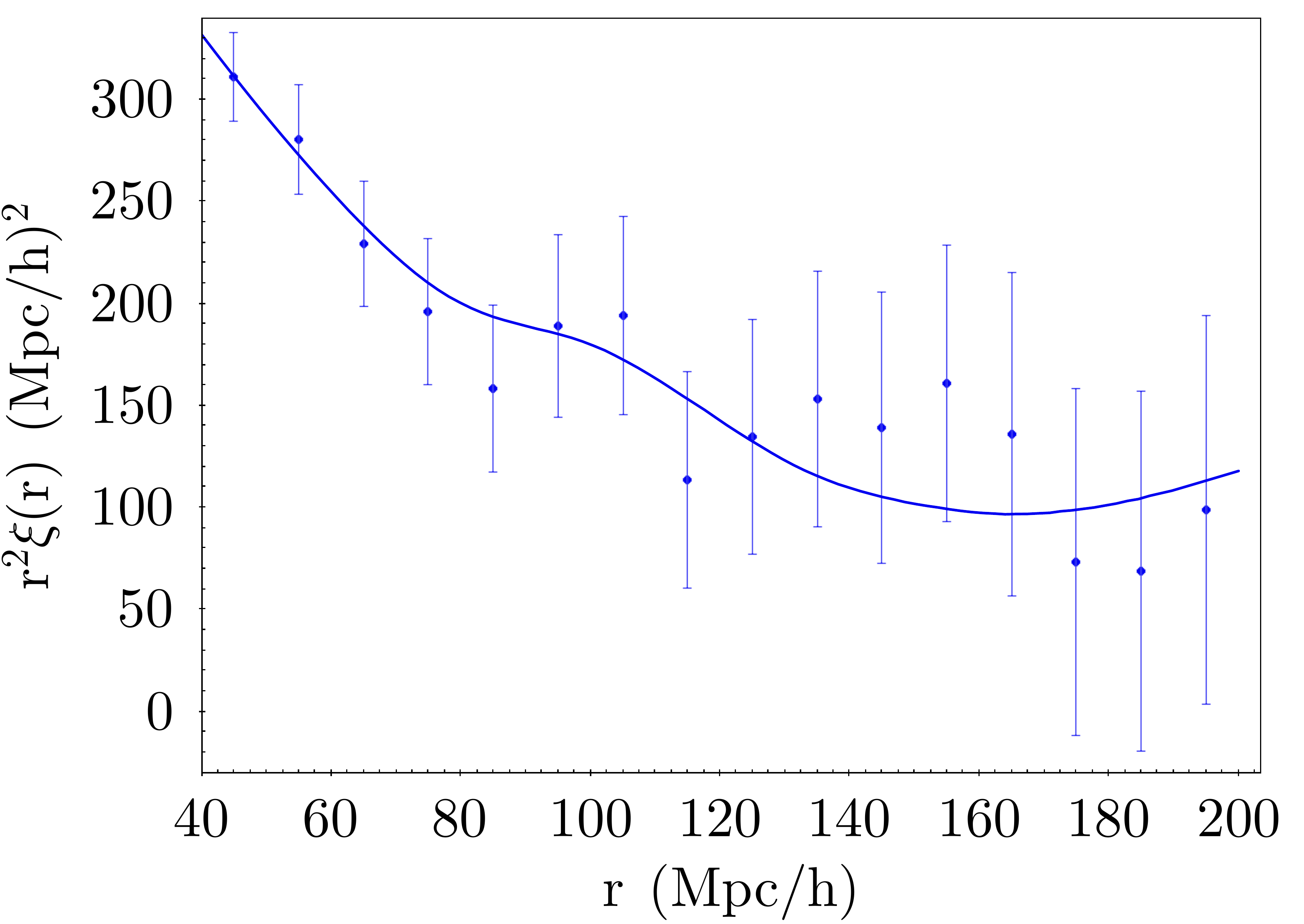} 
   \put(-130,+40){$\alpha = 1.07$}
   \put(-130,30){$\sigma_\alpha =0.14$ }}
 \subfloat
   {\includegraphics[width=0.7\columnwidth]{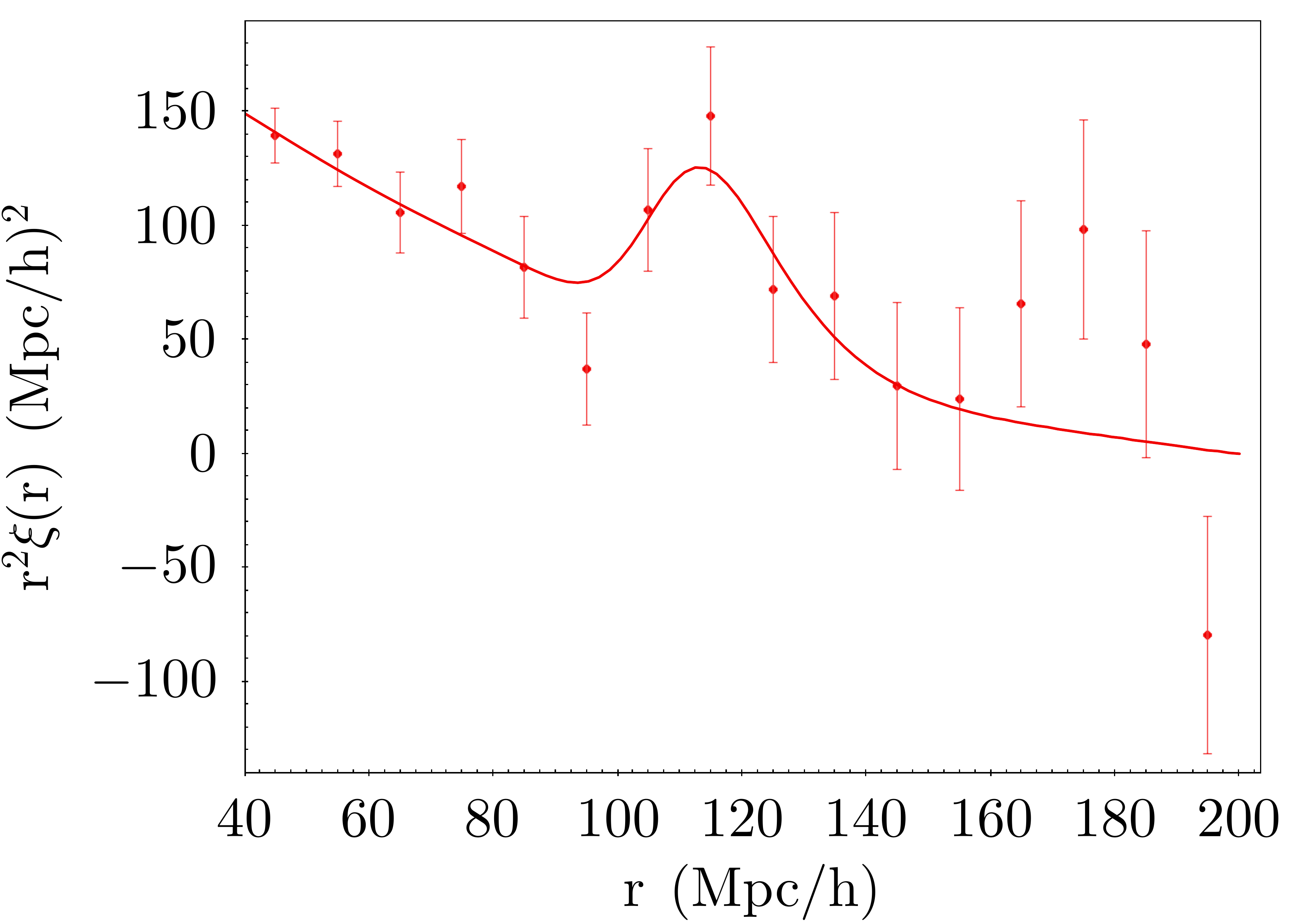}
   \put(-130,+40){$\alpha = 0.98$}
   \put(-130,30){$\sigma_\alpha = 0.03$ }} 
   {\includegraphics[width=0.7\columnwidth]{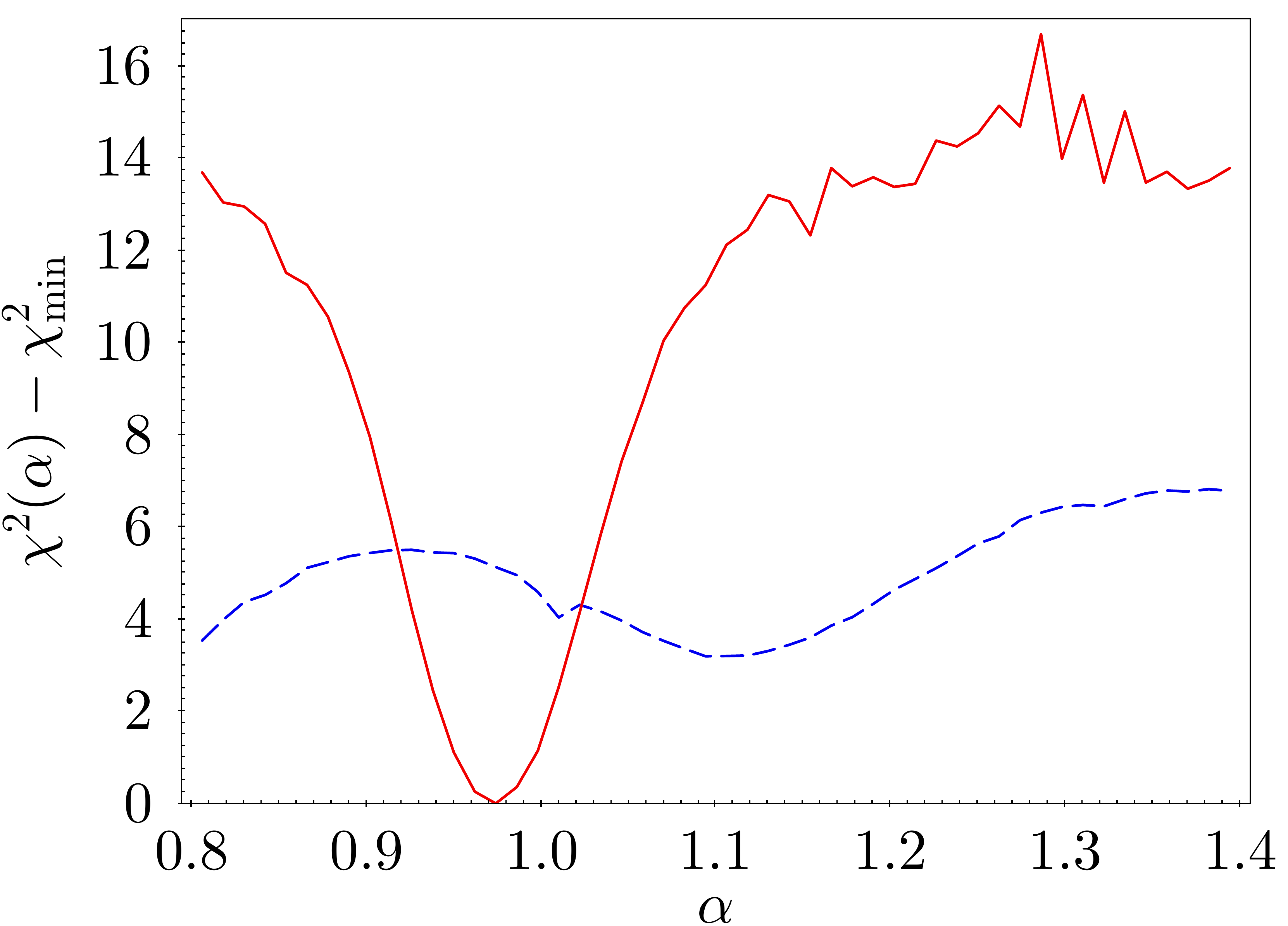}
    \put(-55,30){$\chi^2_\mathrm{min}=23.7$}}\\
 \subfloat
   {\includegraphics[width=0.7\columnwidth]{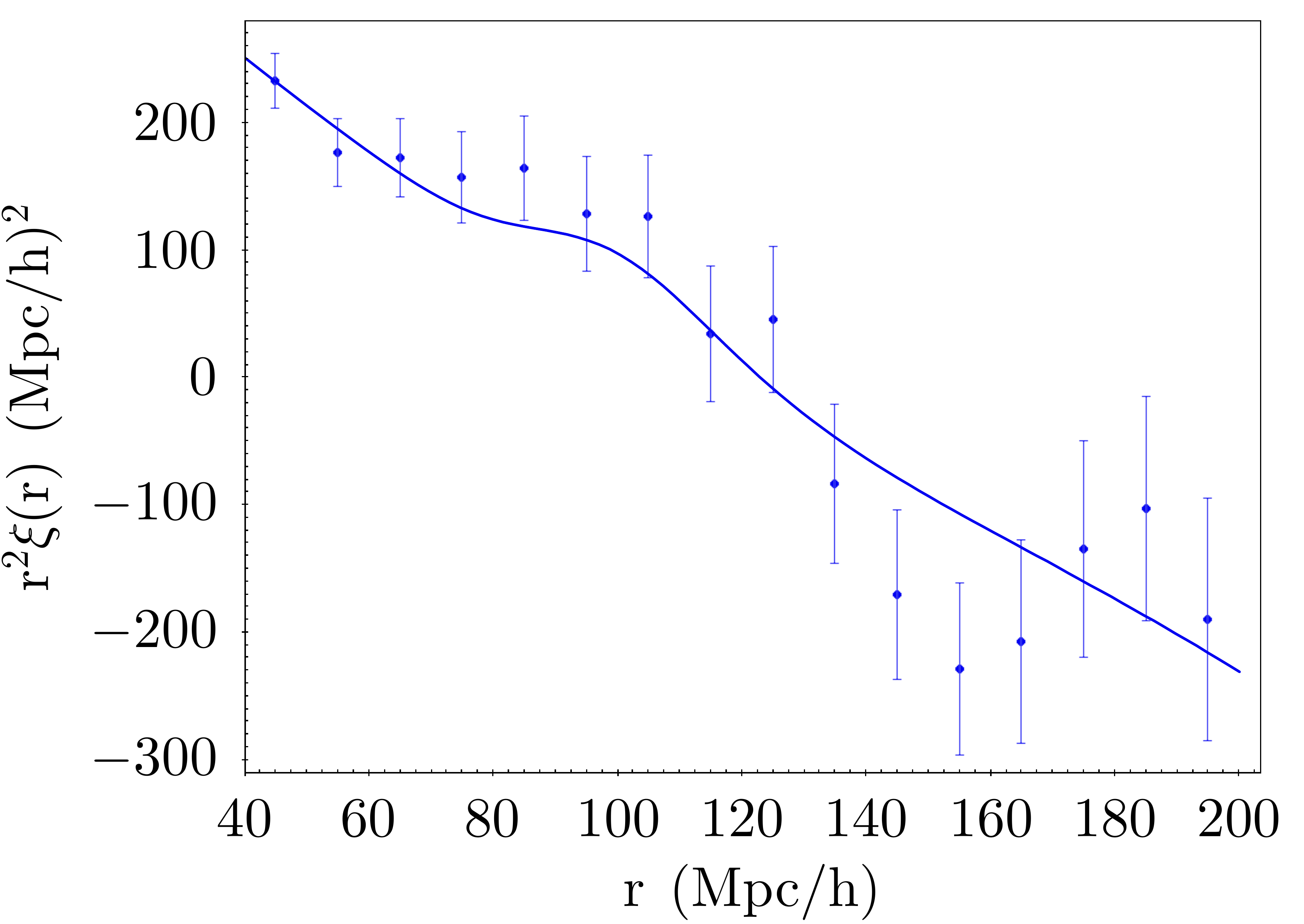} 
   \put(-130,+40){$\alpha = 1.13$}
   \put(-130,30){$\sigma_\alpha = 0.12$ }} 
\subfloat
   {\includegraphics[width=0.7\columnwidth]{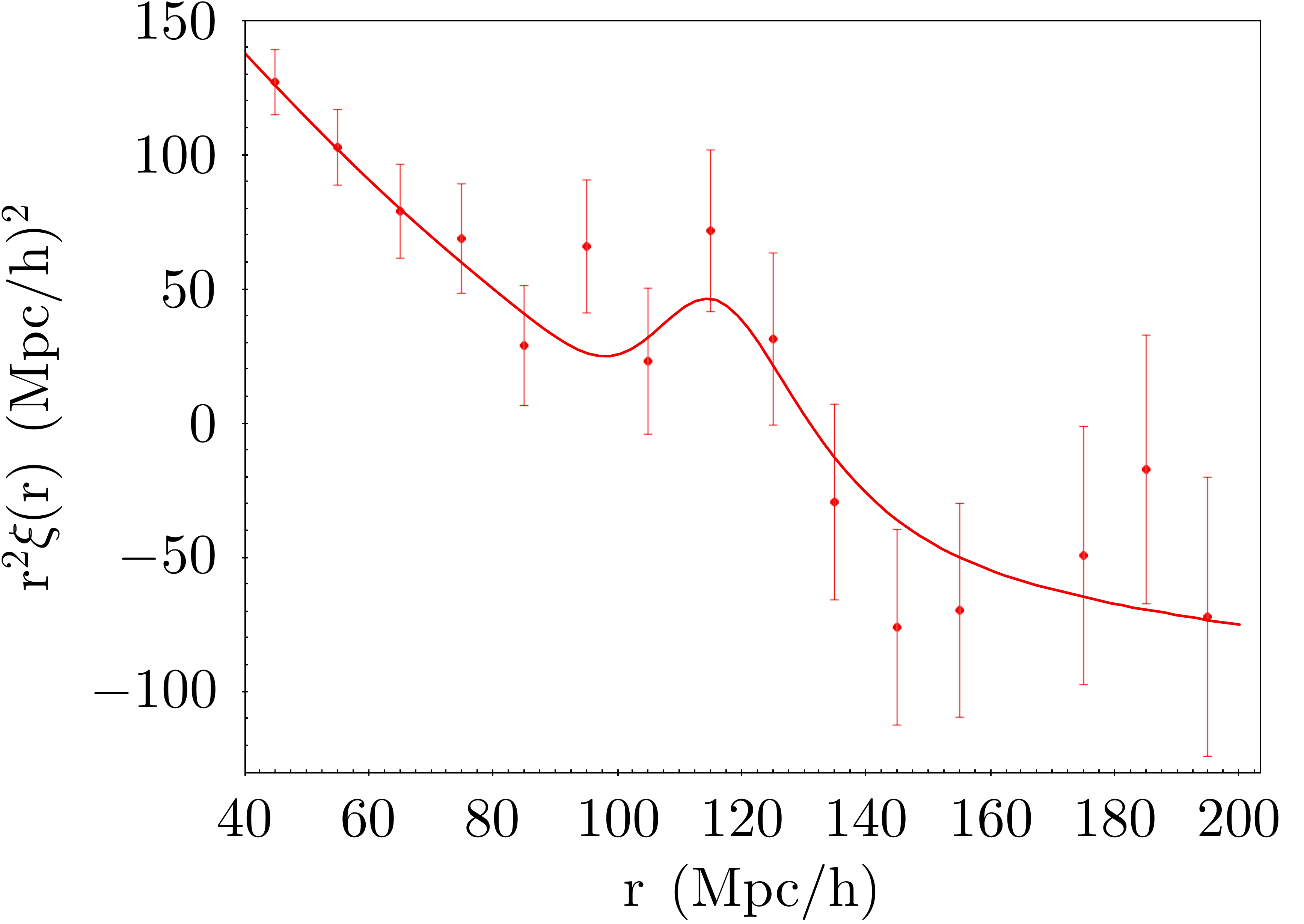}
   \put(-130,+40){$\alpha =0.97 $}
   \put(-130,30){$\sigma_\alpha =0.07$ }} 
\subfloat
   {\includegraphics[width=0.7\columnwidth]{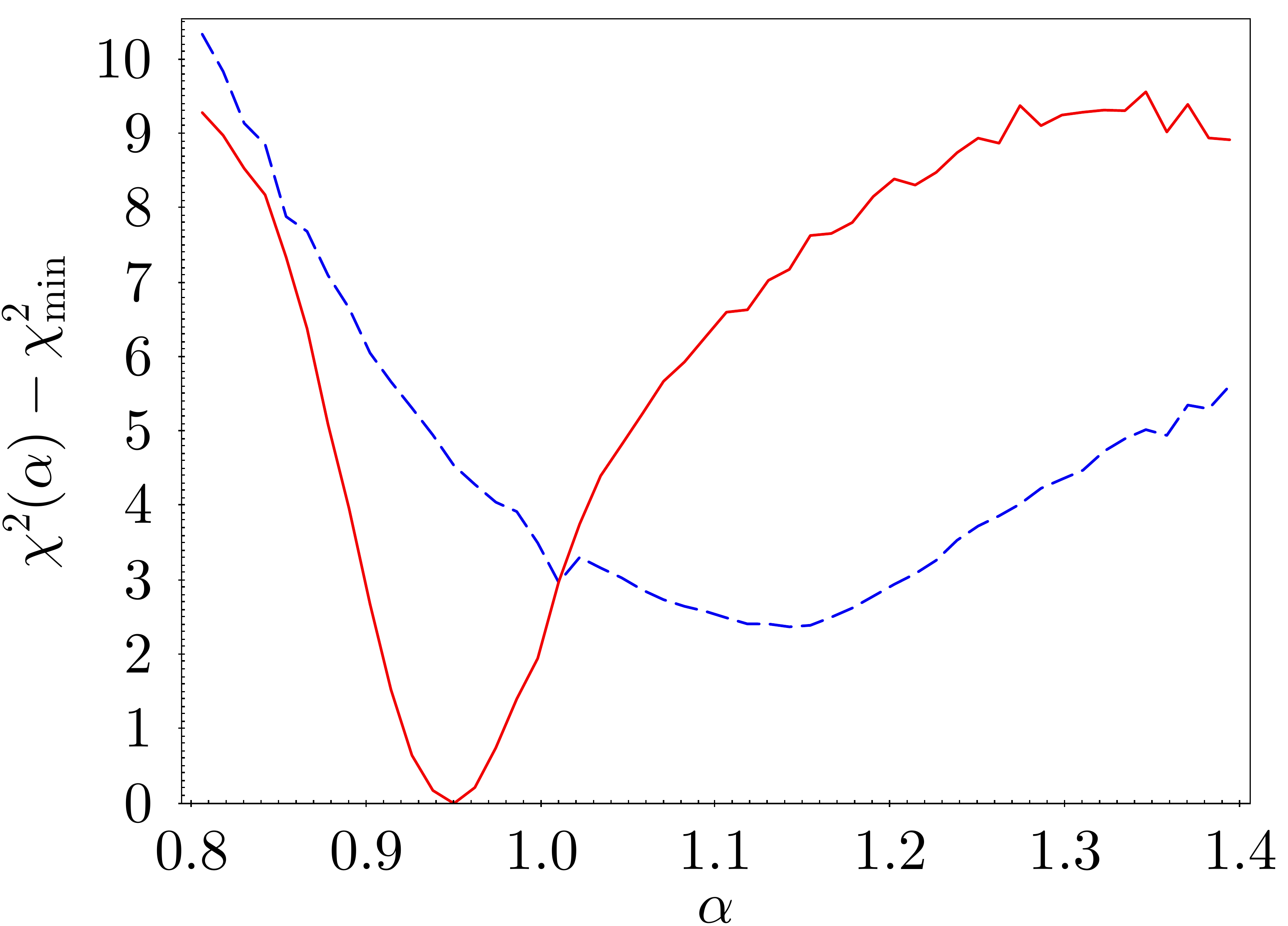}
   \put(-55,30){$\chi^2_\mathrm{min}=11.6$}}\\
\subfloat
   {\includegraphics[width=0.7\columnwidth]{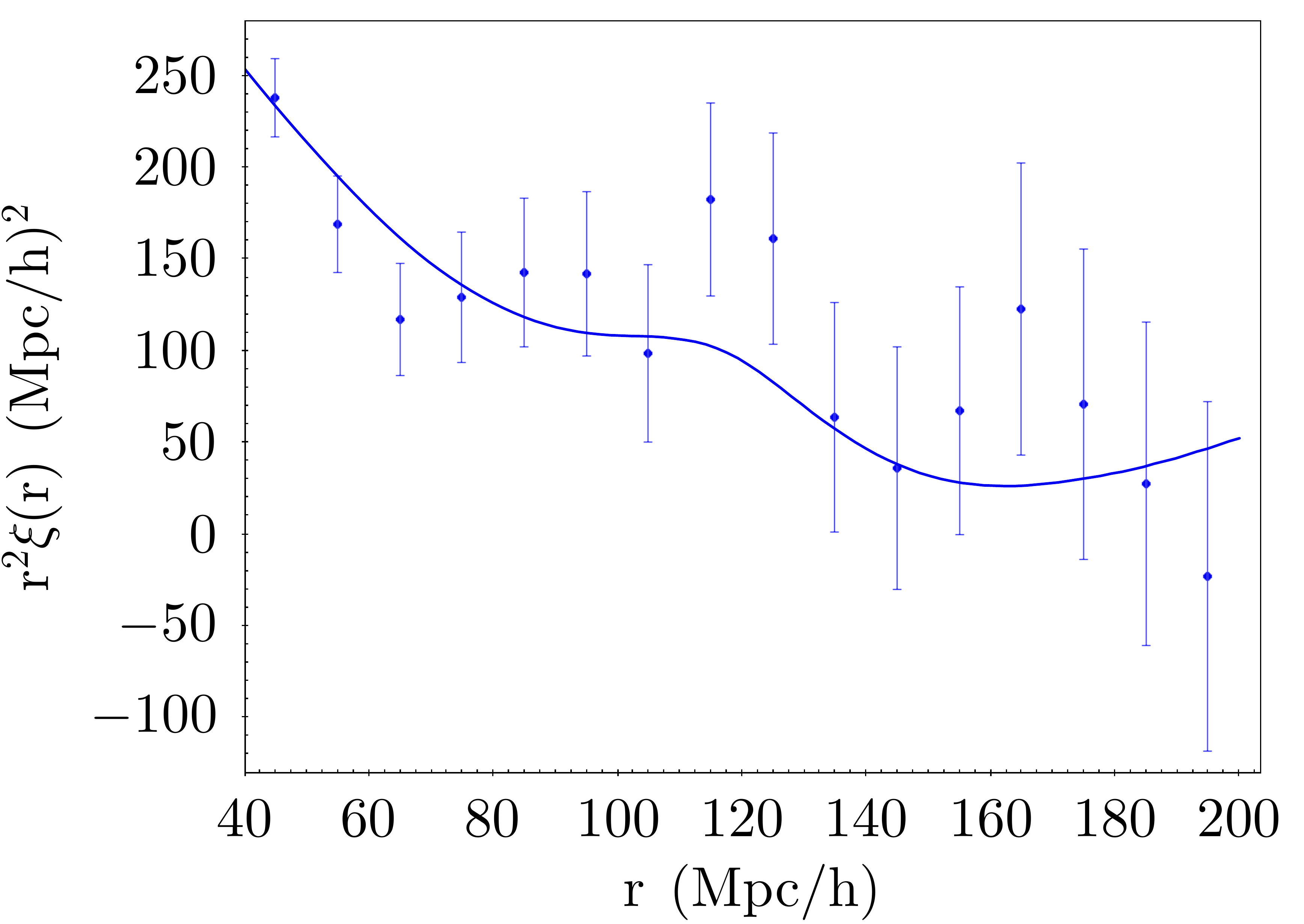} 
   \put(-130,+40){$\alpha = 1.06$}
   \put(-130,30){$\sigma_\alpha = 0.15$ }} 
\subfloat
   {\includegraphics[width=0.7\columnwidth]{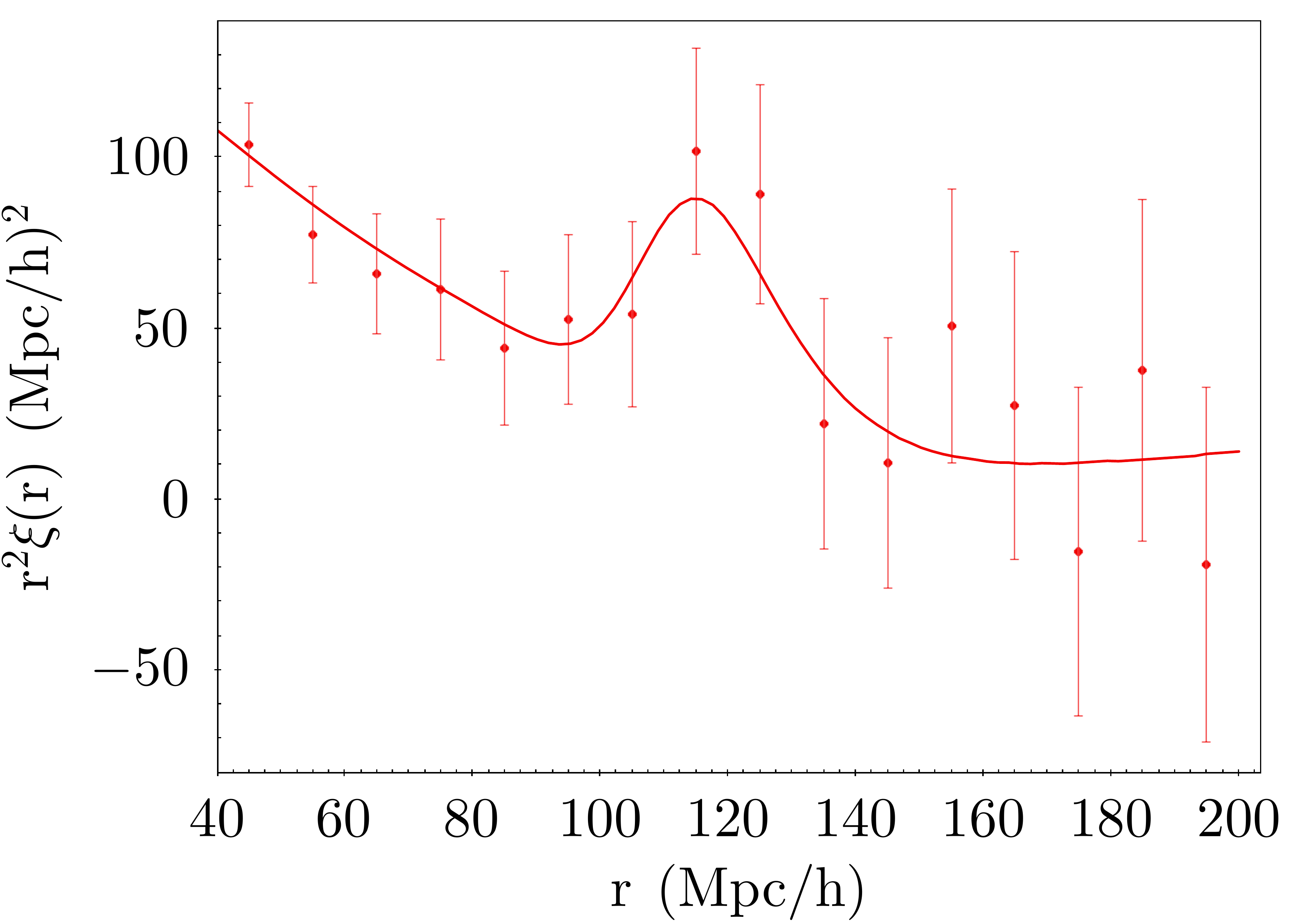}
   \put(-130,+40){$\alpha = 0.97$}
   \put(-130,30){$\sigma_\alpha = 0.05$ }} 
\subfloat
   {\includegraphics[width=0.69\columnwidth]{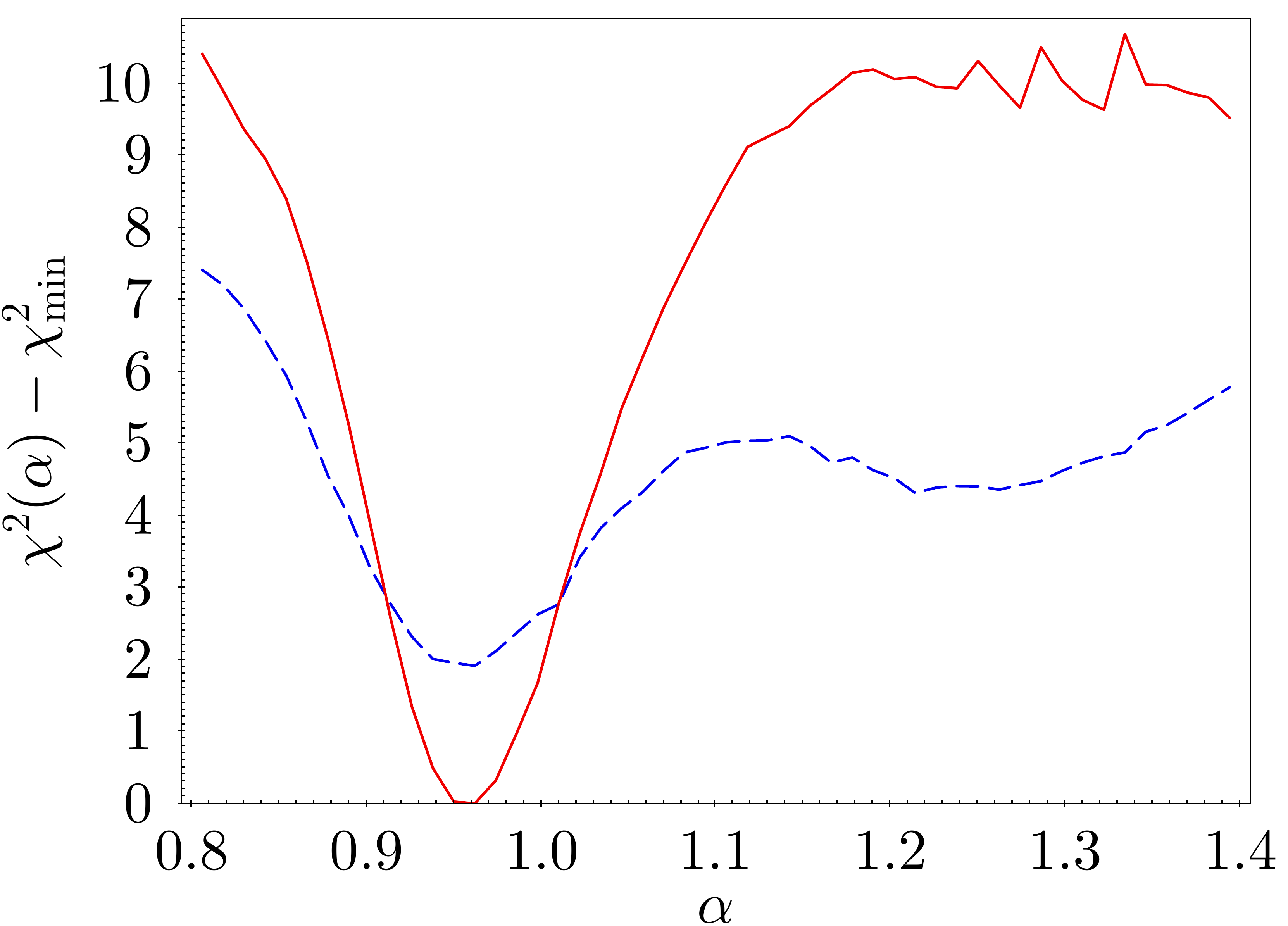}
    \put(-55,30){$\chi^2_\mathrm{min}=26.2$}}\\
\caption{Fit results from the fiducial model for the type-II anomalous samples. {\it Left column:} Monopole of the two-point correlation function before reconstruction; the acoustic peak is not visible. {\it Middle column:} Correlation function after reconstruction; the peak is now clearly visible and the fit returns an unbiased value of $\alpha$. {\it Right column:} $\Delta\chi^2(\alpha)=\chi^2(\alpha)-\chi^2_\mathrm{min}$ before reconstruction (blue-dashed line) and after reconstruction (red line); the minimum of $\chi^2(\alpha)$ curve that pre-reconstruction was either unclear or not unique becomes well-defined post-reconstruction.}
\label{fig:Unl_peak_rspace}
\end{figure*}

For illustrative purposes, two representative examples of type-I anomalous samples are shown in Figure~\ref{fig:Unl_shift_rspace}. The correct shift of the BAO peak in the two-point correlation function from its incorrect position pre-reconstruction (left column) to the right position around $\sim 110h^{-1}$Mpc post-reconstruction (central column) is obtained, and consequently the correct value of $\alpha$ is measured. Also, the precision of the $\alpha$ estimate increases after the reconstruction. The shift of the minimum of the $\chi^2$ (right column) clearly illustrates and quantifies the de-biasing effect of the reconstruction. Analogously for type-II anomalous samples, three typical examples are shown in Figure~\ref{fig:Unl_peak_rspace}. Here the eFAM algorithm displays its remarkable ability to sharpen and shift the minimum, and consequently to significantly increase the statistical significance of the BAO peak, the ability to reconstruct it at its expected position, and to improve the precision of the measured $\alpha$ value. The sharpening of the BAO peak is paralleled by the decrease of the $\chi^2$ minimum, which was either not present or not unique before the reconstruction.

\subsection{RSD modelling: consistency tests}\label{subsec:vvcomparison}

\begin{figure}
\centering
\includegraphics[width=0.9\columnwidth]{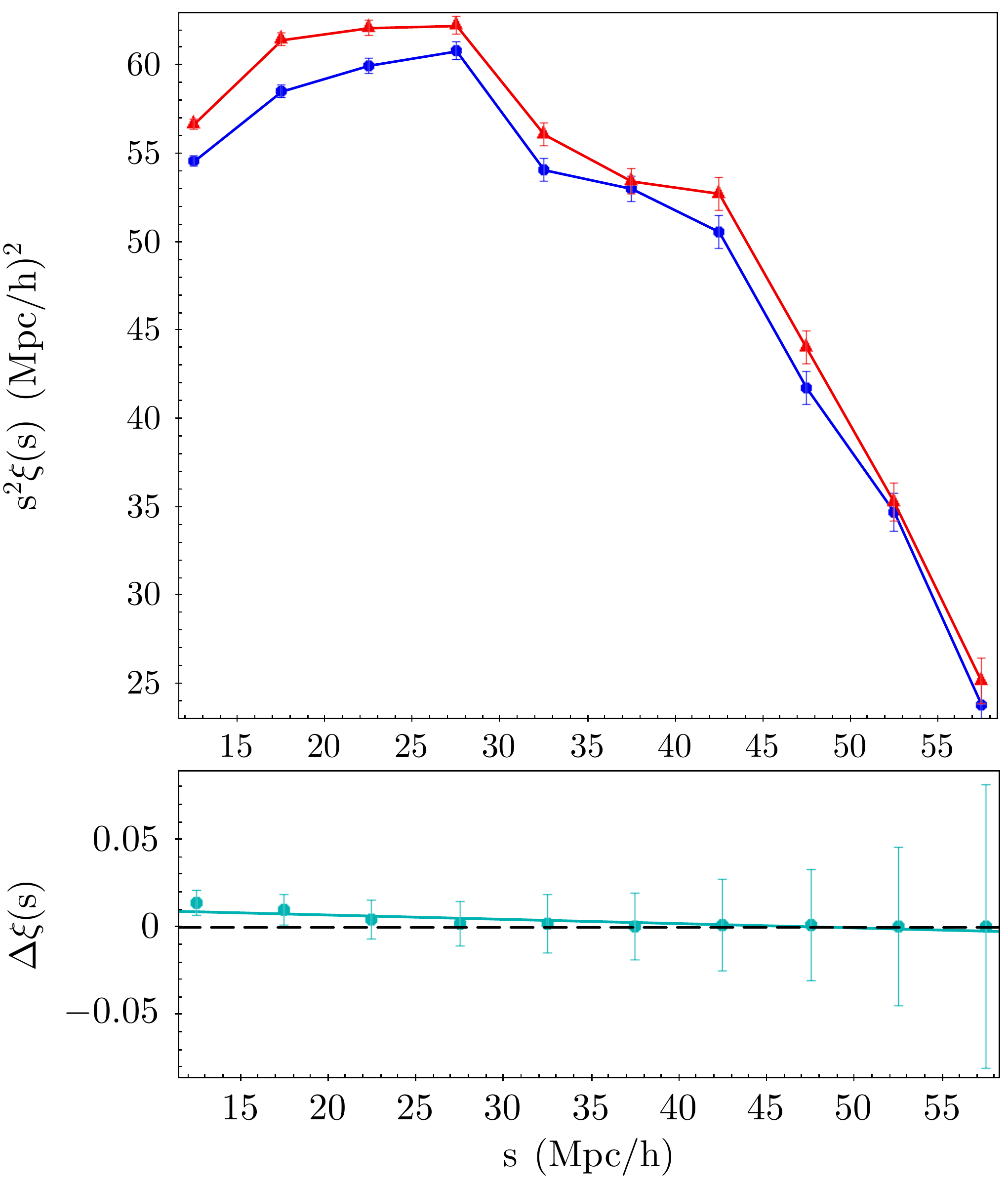}
\caption{Accuracy test of eFAM$_{10}$ for the clustering statistics in redshift-space. \emph{Top:} Rescaled monopole of the two-point correlation function computed from the haloes with known velocities (blu lines, circles) and with eFAM peculiar velocities (red line, triangles). \emph{Bottom:} Residuals of the correlation function. Assigning eFAM velocities to build the halo catalogues in redshift-space results in an overestimation of the amplitudes of the correlation function, which decreases with increasing separation.}
\label{fig:Monopole_test}
\end{figure}

The velocities of \textsc{deus-fur} haloes are not supplied. Since these are necessary to set the initial condition of the eFAM reconstruction in redshift-space, we use the velocity predicted by eFAM itself. There is of course a certain degree of circularity in this procedure that may, in principle, artificially increase the accuracy of the reconstruction itself. To investigate this issue we run a specific test in which we considered a set of halos extracted from the same \textsc{deus} simulation described in Section \ref{subsec:tidal} using a standard FoF algorithm that returns mass, position and centre of mass velocity of each object.vIn the test we run two FAM reconstructions: one in which FoF velocities are used to set the initial conditions and the other in which we use FAM velocities. We then compared the monopole of the two-point correlation function computed from the haloes with known velocities, with the one obtained using the output of the reconstruction; see top panel of Figure~\ref{fig:Monopole_test}. At small separation the strong two-point correlation between the velocities of particles increases the amplitude of the correlation function in redshift-space, $\xi(s)$. This effect is artificially magnified when eFAM velocities are considered. However, this effect decreases with increasing relative separation when the velocities of the pairs become less correlated, asymptotically tending to zero at large scales; see bottom panel of Figure~\ref{fig:Monopole_test}. We therefore expect to obtain unbiased correlation functions at the BAO scale, which supports the robustness of our procedure.
It is worth noticing that the number density of halos in this test is approximately 100 times higher than in the BAO reconstructions, making this a very demanding consistency test. Moreover, no additional smoothing has been applied to mitigate nonlinear effects and the same geometry and treatment of the tidal field as in the BAO tests has been adopted.

\begin{figure*}
\centering
\includegraphics[width=0.95\columnwidth]{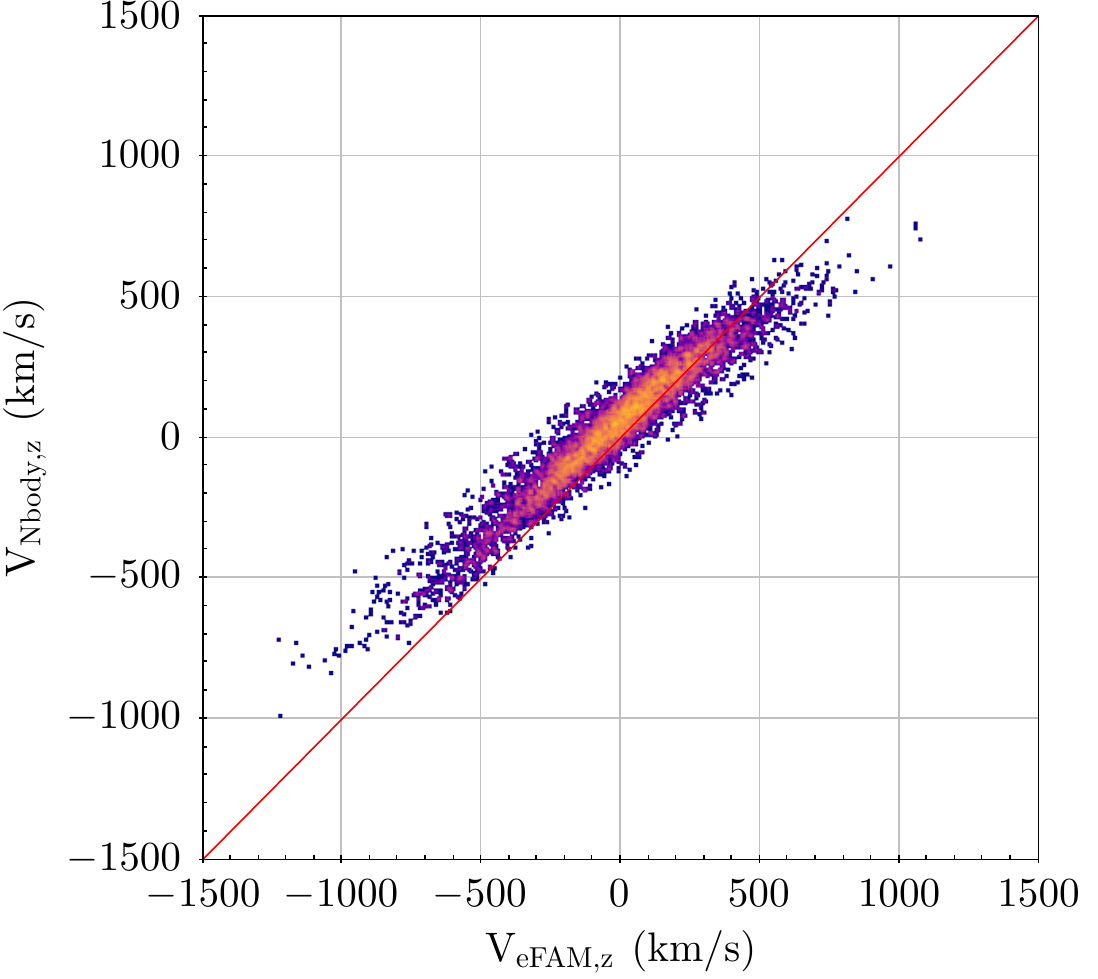}
\includegraphics[width=0.965\columnwidth]{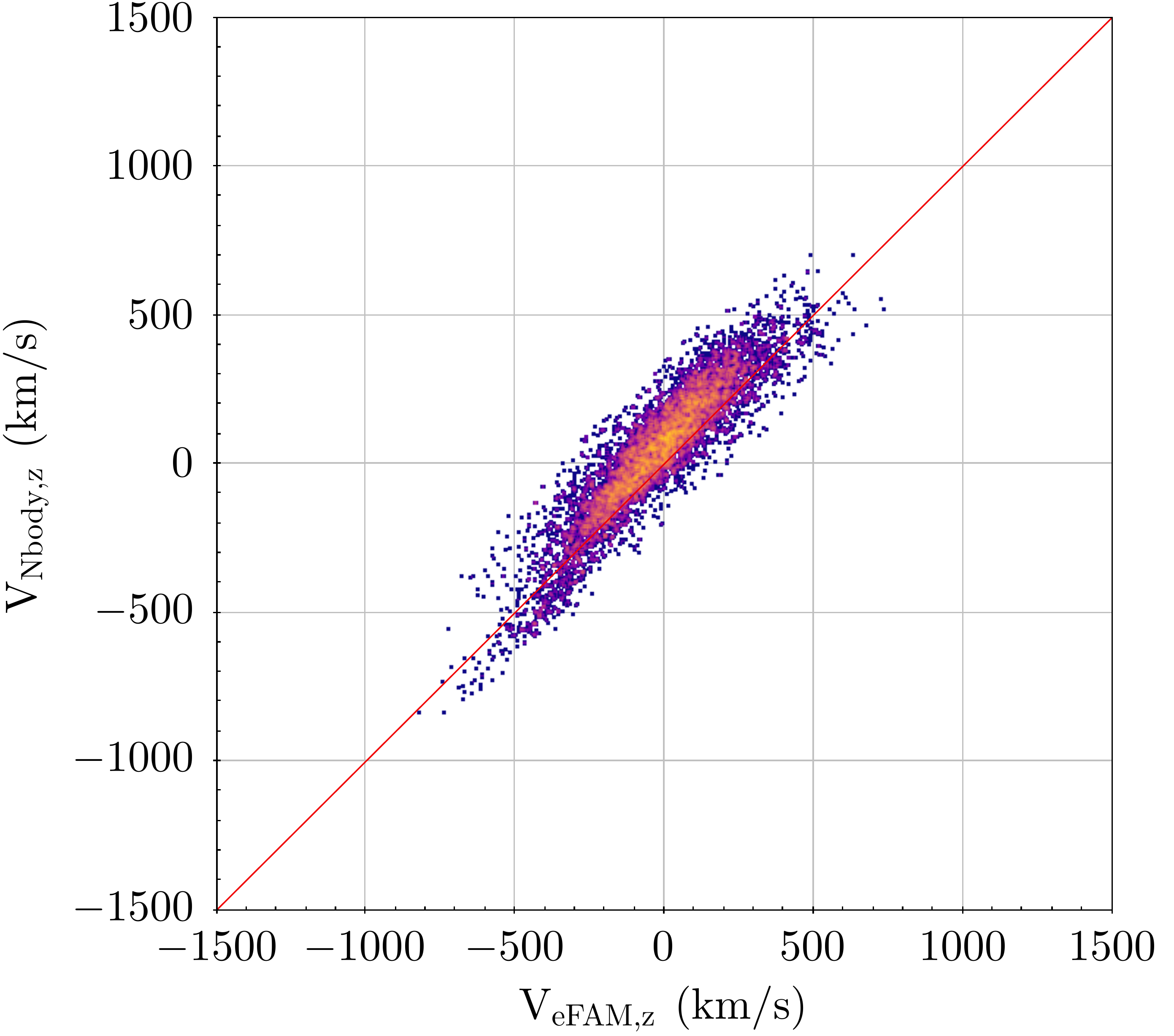}
\put(-325,55){$m =0.735\pm0.003$}
\put(-325,45){$q=(-10\pm1)$~km\,s$^{-1}$}
\put(-325,35){$\sigma =61$~km\,s$^{-1}$}
\put(-87,55){$m = 0.920 \pm 0.005 $}
\put(-87,45){$q = (8 \pm 2)$~km\,s$^{-1}$}
\put(-87,35){$\sigma = 81$~km\,s$^{-1}$}
\caption{Accuracy tests of eFAM$_{10}$ in real-space (left) and redshift-space (right). Reconstructed versus true peculiar velocities of haloes at $z=0$ for one Cartesian component (results are similar for other components). A perfect reconstruction would give a linear regression $V_\mathrm{eFAM}=mV_\mathrm{Nbody}+q$ with slope $m=1$ (solid line), no residual bulk velocity ($q=0$), and no scatter. The reconstructed peculiar velocities in real-space are slightly overestimated though well-correlated with the true ones. In redshift-space the reconstructed peculiar velocities are definitely less biased but more scattered as expected.}
\label{fig:Scatter_plot_RSspace}
\end{figure*}
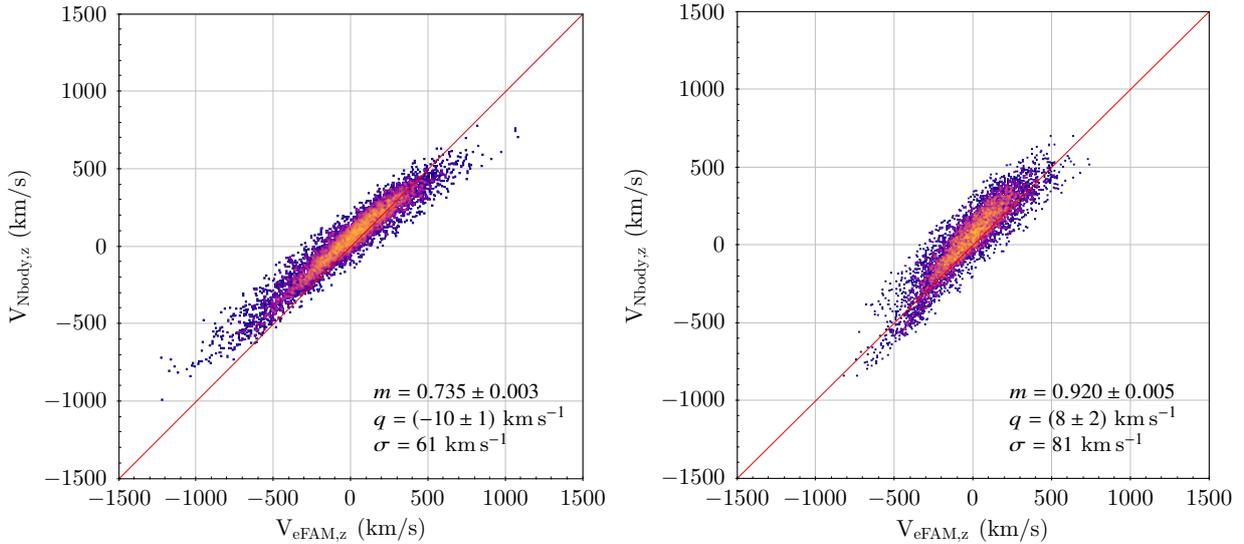

A more demanding test consists in comparing halo-by-halo the Cartesian components of the true velocities $\mathbf{v}_{i,\mathrm{Nbody}}$ with those reconstructed by eFAM$_{10}$, $\mathbf{v}_{i,\mathrm{eFAM}}$. In real-space (Figure~\ref{fig:Scatter_plot_RSspace}, left panel), regardless of the Cartesian component, the reconstructed velocities are typically overestimated by $20-25$ percent irrespective of the extension of the outskirt accounting for the tidal field, with small and constant dispersion for velocities $\lesssim 1000$~km\,s$^{-1}$. This is a well-known effect, already pointed out by \citet{BranchiniEldarNusser2002} due to the biased density field that we use to compute the gravitational potential. While in linear theory this overestimation can be approximately undone by normalising the reconstructed velocities by a factor $1/b$, with $b$ the effective halo bias at $z=0$, the correction for FAM velocities is more complicated since this method goes beyond the linear theory. To correct for this effect, we weighted each halo mass by the number of dark matter particles within it. Though not accurate, this recipe does provide an approximate correction accounting for the haloes as biased tracers of the mass distribution. Besides, note that $N$-body velocities contain incoherent non-linear components that are not captured by the FAM reconstructed velocities. The small amplitude of the offset, reduced to about 10~km\,s$^{-1}$ after considering an external buffer of thickness $200 h^{-1}$Mpc, indicates a not exact though largely sufficient modelling of the negligible bulk-flow offset. It is worth to note that an overestimation of velocities by $\sim20$ percent in amplitude shall result in an error on the redshift coordinate of about $\Delta s= 0.2 v_\mathrm{Nbody}/c\sim 2.7\times 10^{-4}$ for the typical value $v_\mathrm{Nbody}=400$~km\,s$^{-1}$, which is less than the usual error on the spectroscopic measurement of redshift.

In redshift-space (Figure~\ref{fig:Scatter_plot_RSspace}, right panel) the overestimation of reconstructed velocities is reduced to $\sim8$ percent but slightly more scattered, with similar bulk-flow as in real-space. This is an effect of the Fingers-of-God, which act as a natural smoothing of the density field dumping the amplitudes of peculiar velocities. Accordingly, smaller peculiar velocities allow a reconstruction pushed at earlier time, from $z\sim 7$ in real-space to $z\sim 40$ in redshift-space using eFAM$_{10}$.

\begin{figure*}
\centering
\subfloat
   {\includegraphics[width=0.672\columnwidth]{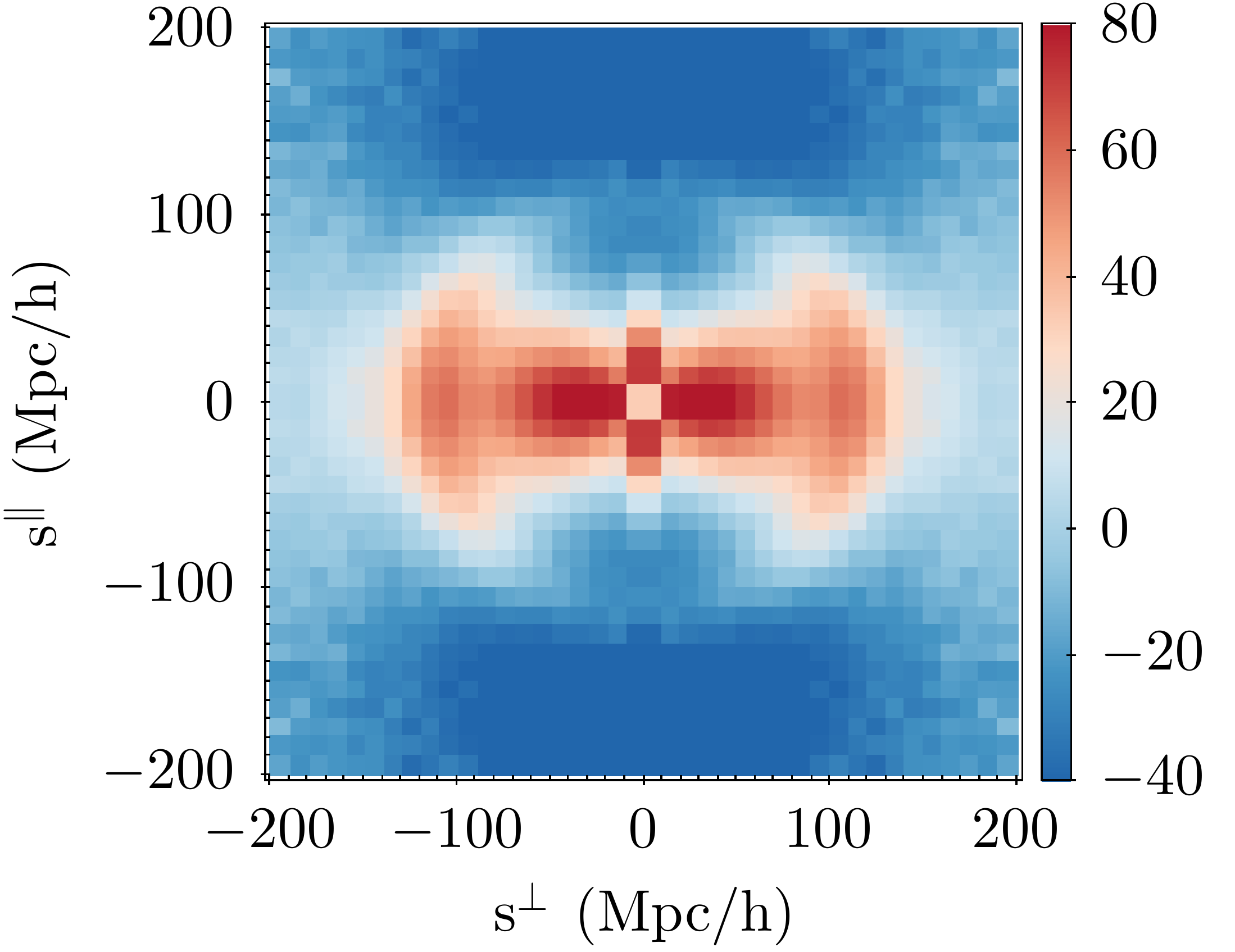} } \quad
\subfloat
   {\includegraphics[width=0.672\columnwidth]{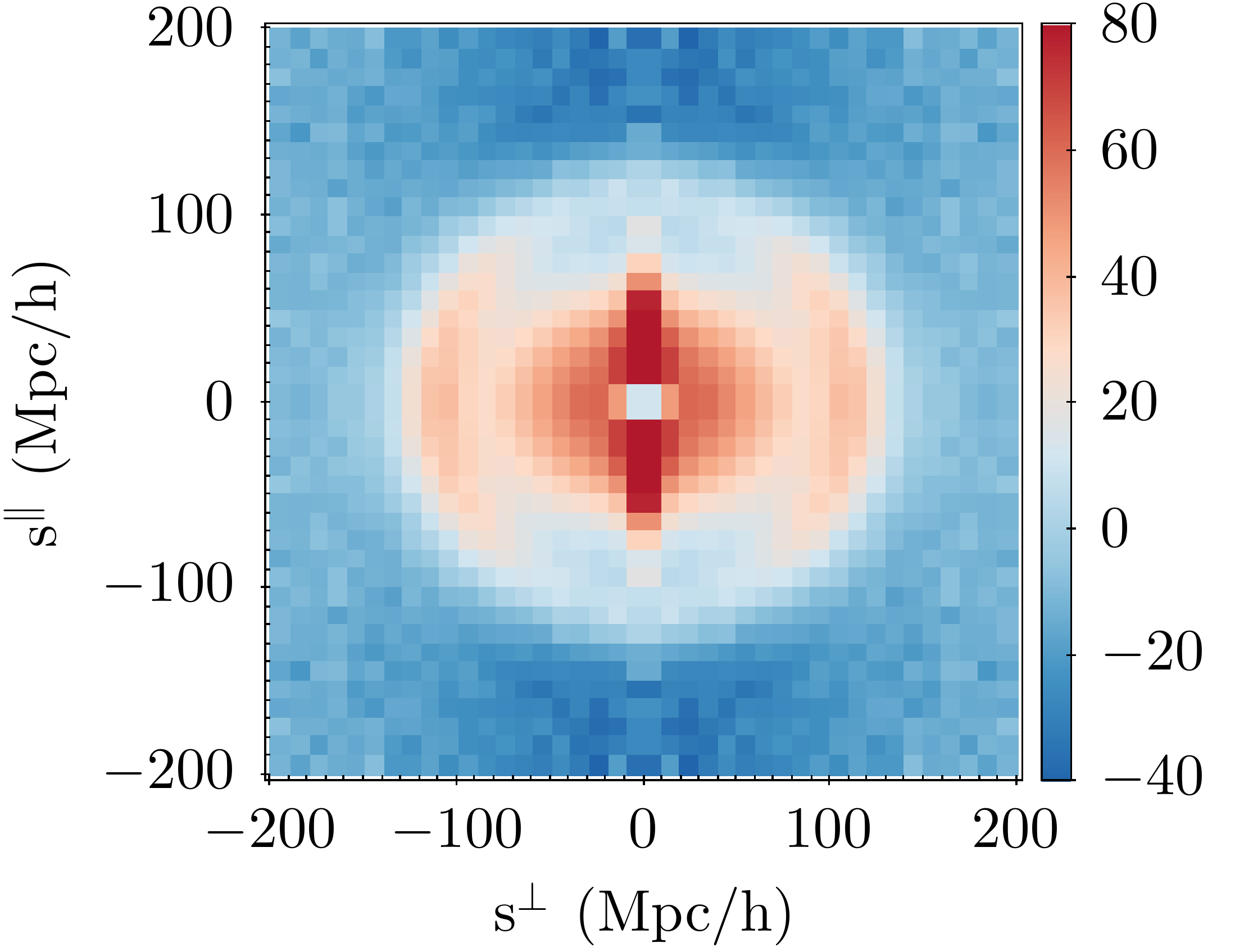}} \quad
\subfloat
   {\includegraphics[width=0.672\columnwidth]{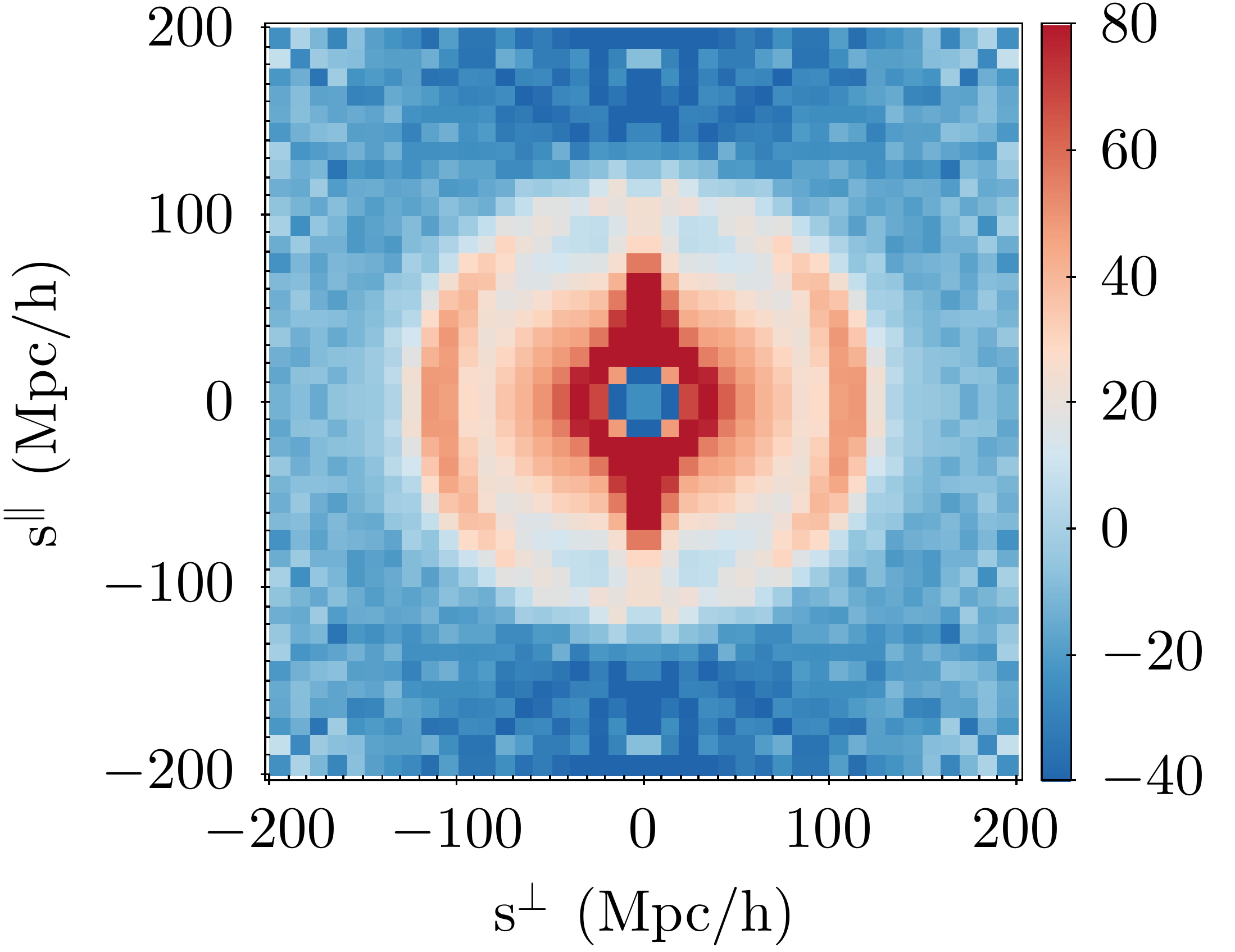}}
   \caption{Rescaled two-point correlation function in redshift-space, averaged over the 512 mocks, as function of the transverse ($s^\perp$) and line-of-sight ($s^\parallel$) separation.
   {\it Left panel:} Before reconstruction, from objects at $z=0$; the isotropy of the acoustic feature, which should be visible as a ring in the ($s^\parallel$, $s^\perp$) plane, is broken by RSD.
      {\it Middle panel:} Correlation function after correcting the density field for the RSD, again at the observed redshift $z=0$; the isotropy of the acoustic feature is almost completely restored.
   {\it Right panel:} Correlation function after reconstruction at $z=33.6$; the BAO feature is sharper and symmetric, indicating the quality of the reconstruction.}
\label{fig:xi_r_pi}
\end{figure*}

\subsection{Monopole, quadrupole, and anisotropic correlation function in redshift-space}

The attractive feature of the eFAM technique is to recover the peculiar velocities of objects at their observed redshift, allowing for a non-parametric modelling of the RSD.
This is illustrated in Figure~\ref{fig:xi_r_pi}, which shows the density plots of the rescaled anisotropic two-point correlation function $s^2\bar{\xi}(s^\parallel,s^\perp)$ as function of the longitudinal ($s^\parallel$) and transverse ($s^\perp$) components of the separation vector $\mathbf{s}$, averaged over the 512 mocks. If the reconstruction is successful, $s$ represents the cosmological redshift with no peculiar velocity component in it. Before reconstruction (left panel), the isotropy of the correlation function is broken by the RSD, which compresses the BAO ring at $\sim110h^{-1}$Mpc along the line-of-sight and split it into two arcs. This deformation is almost completely removed after correcting the density field in redshift-space by subtracting the longitudinal displacement due to the peculiar velocities, as estimated by eFAM at the same redshift of objects (middle panel).
The BAO ring is further sharpened by reconstructing the density field at higher redshift (right panel).

\begin{figure}
 \includegraphics[width=0.9\columnwidth]{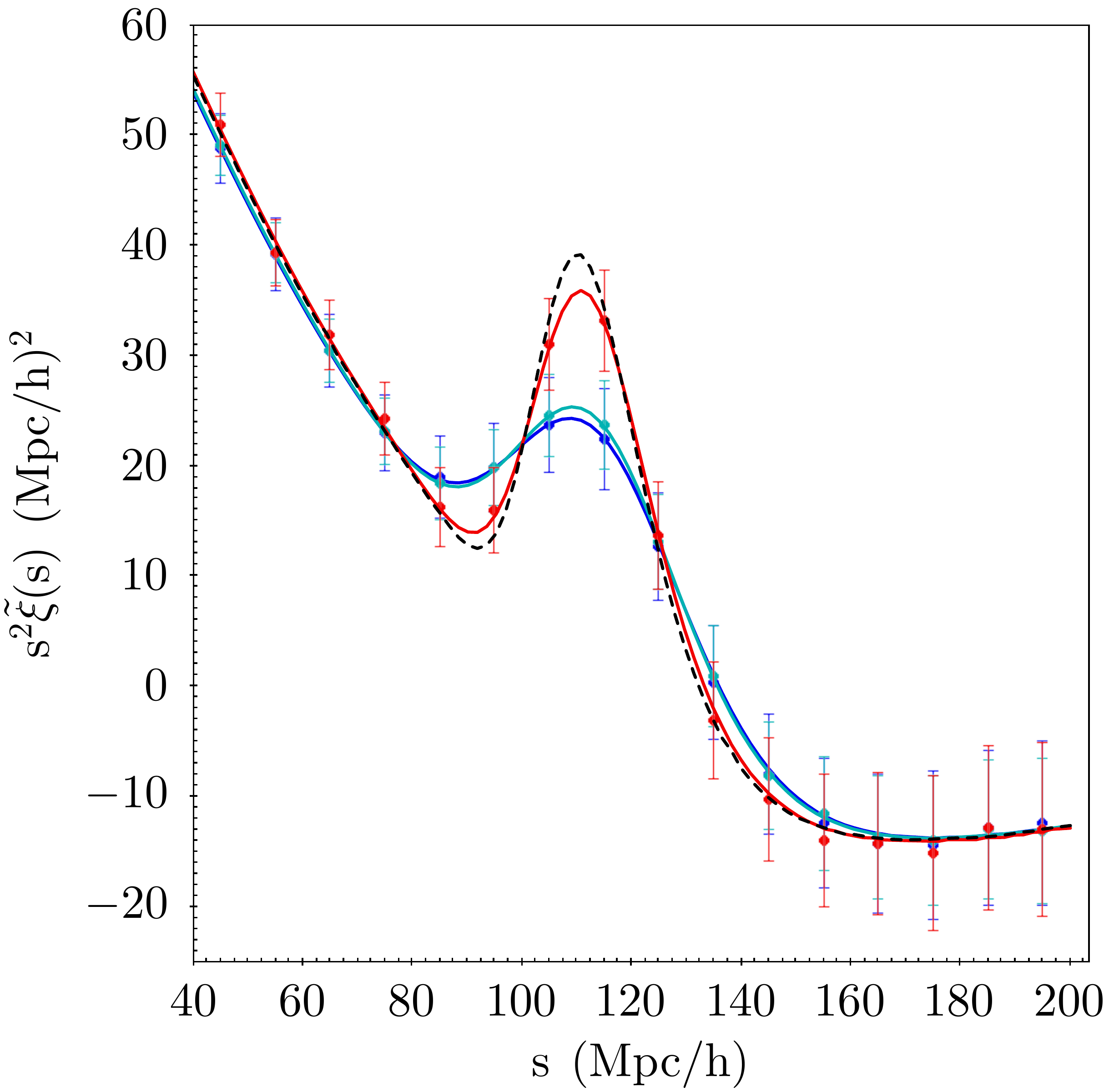}
 \includegraphics[width=0.9\columnwidth]{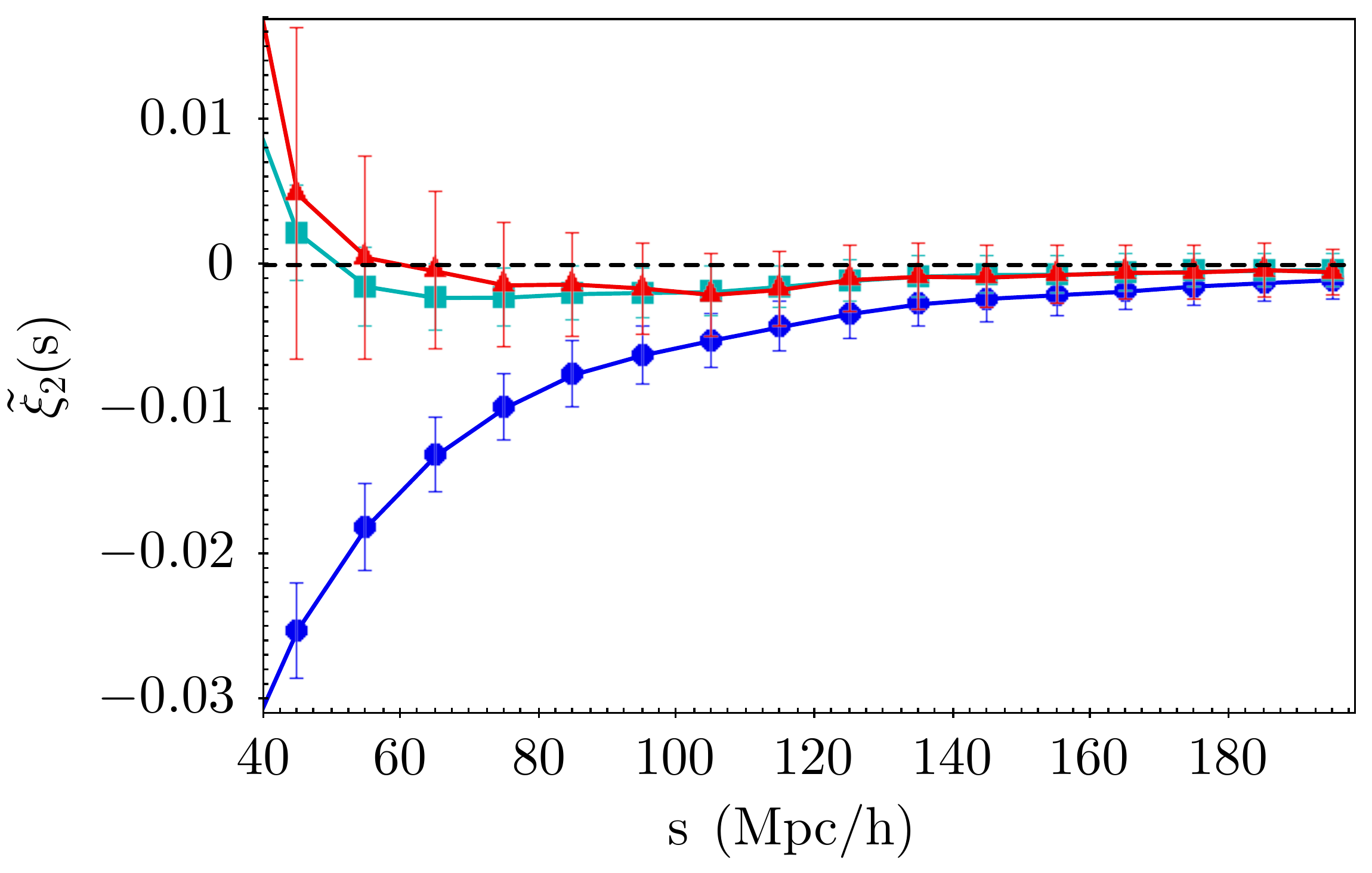}
\caption{Rescaled monopole (top) and quadrupole (bottom) of the two-point correlation function in redshift-space pre- and post-reconstruction by eFAM$_{10}$, averaged over the 512 mocks; error bars from the diagonal of the covariance matrix. Before reconstruction at observed redshift $z=0$ (blue line, circles) the acoustic feature in the monopole is broadened by non-linear evolution and peculiar velocities. The correction for RSD is effective at $z=0$ (green line) as shown in the quadrupole, but the BAO peak is only slightly enhanced. Pushing the reconstruction at the highest redshift possible, $z=33.6$ (red line), the monopole well approximates the linear model (dashed line) and the quadrupole is still consistent with zero, showing the efficiency of the eFAM method in both sharpening the peak and correcting for RSD.}
\label{fig:eFAM_xi0xi2_med_sspace}
\end{figure}

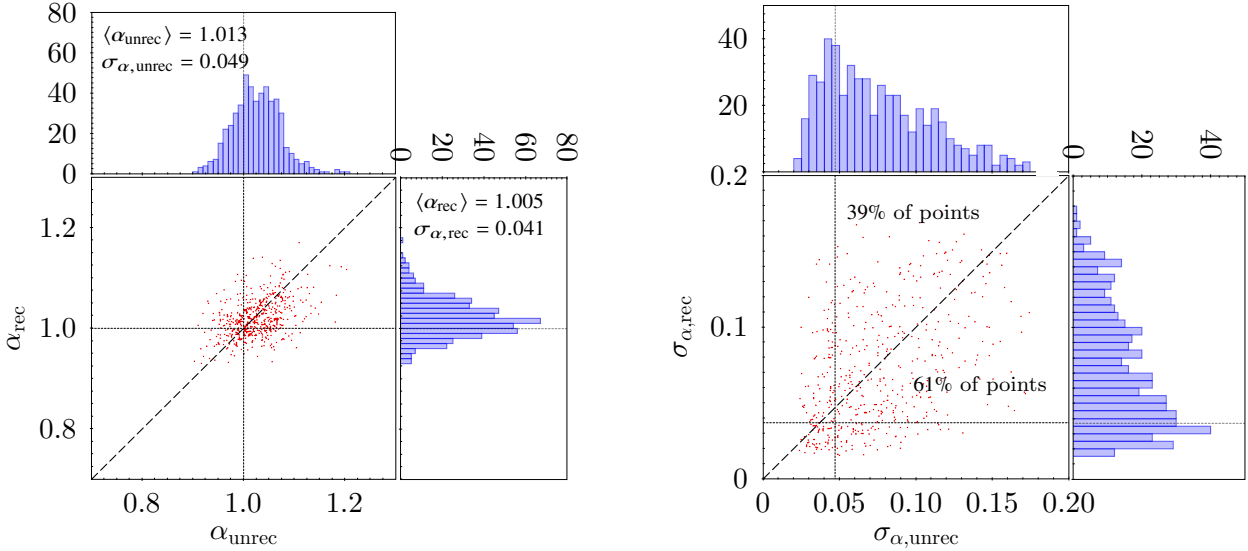
\begin{figure*}
\centering
\subfloat
   {\includegraphics[width=0.9\columnwidth]{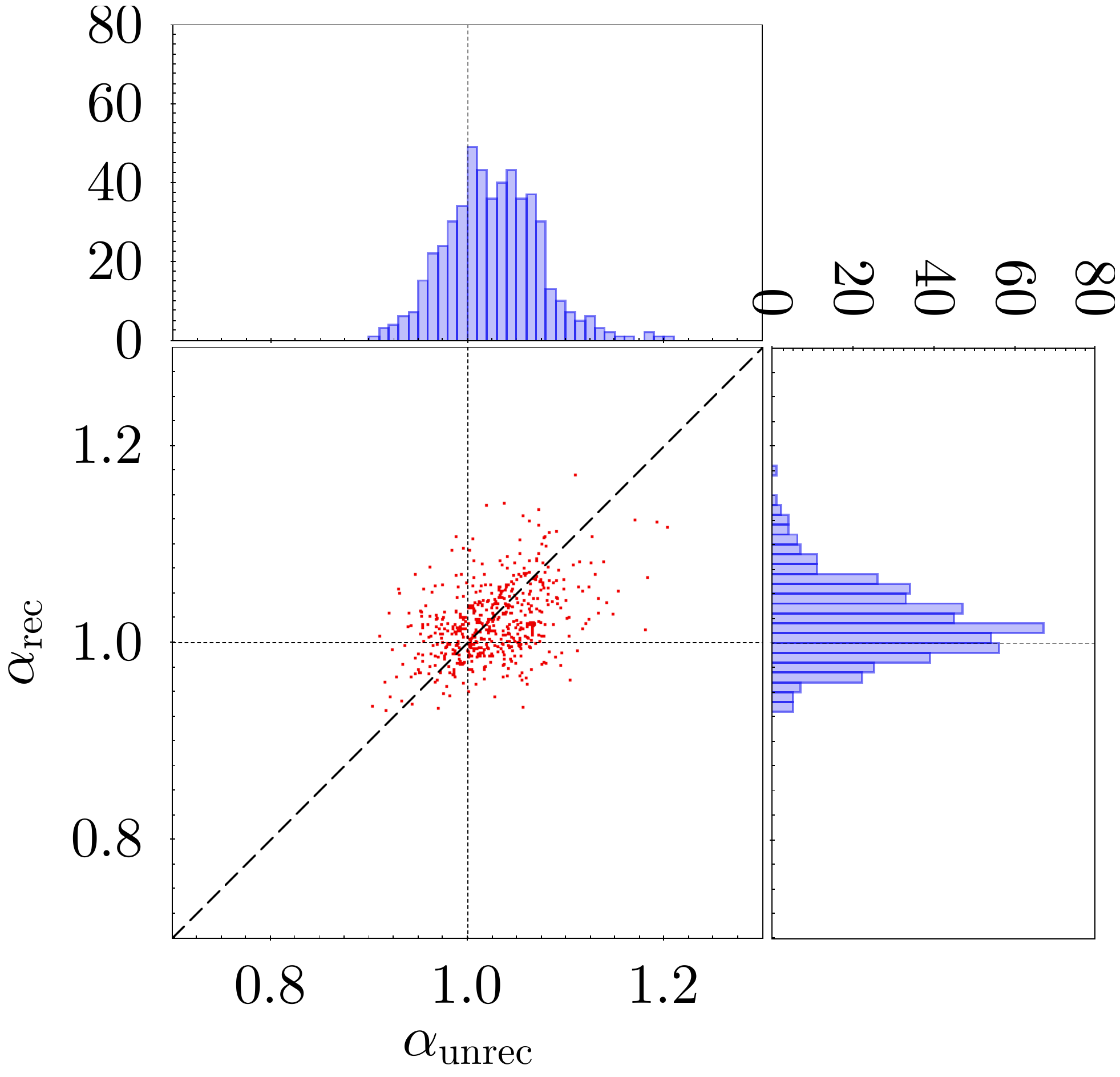}
   \put(-179,192){$\langle\alpha_\mathrm{unrec}\rangle = 1.013$}
   \put(-179,182){$\sigma_{\alpha,\mathrm{unrec}} = 0.049$}
   \put(-62,129){$\langle\alpha_\mathrm{rec}\rangle = 1.005 $}
   \put(-62,119){$\sigma_{\alpha,\mathrm{rec}} = 0.041$}} \quad \quad \quad \quad
\subfloat
   {\includegraphics[width=0.9\columnwidth]{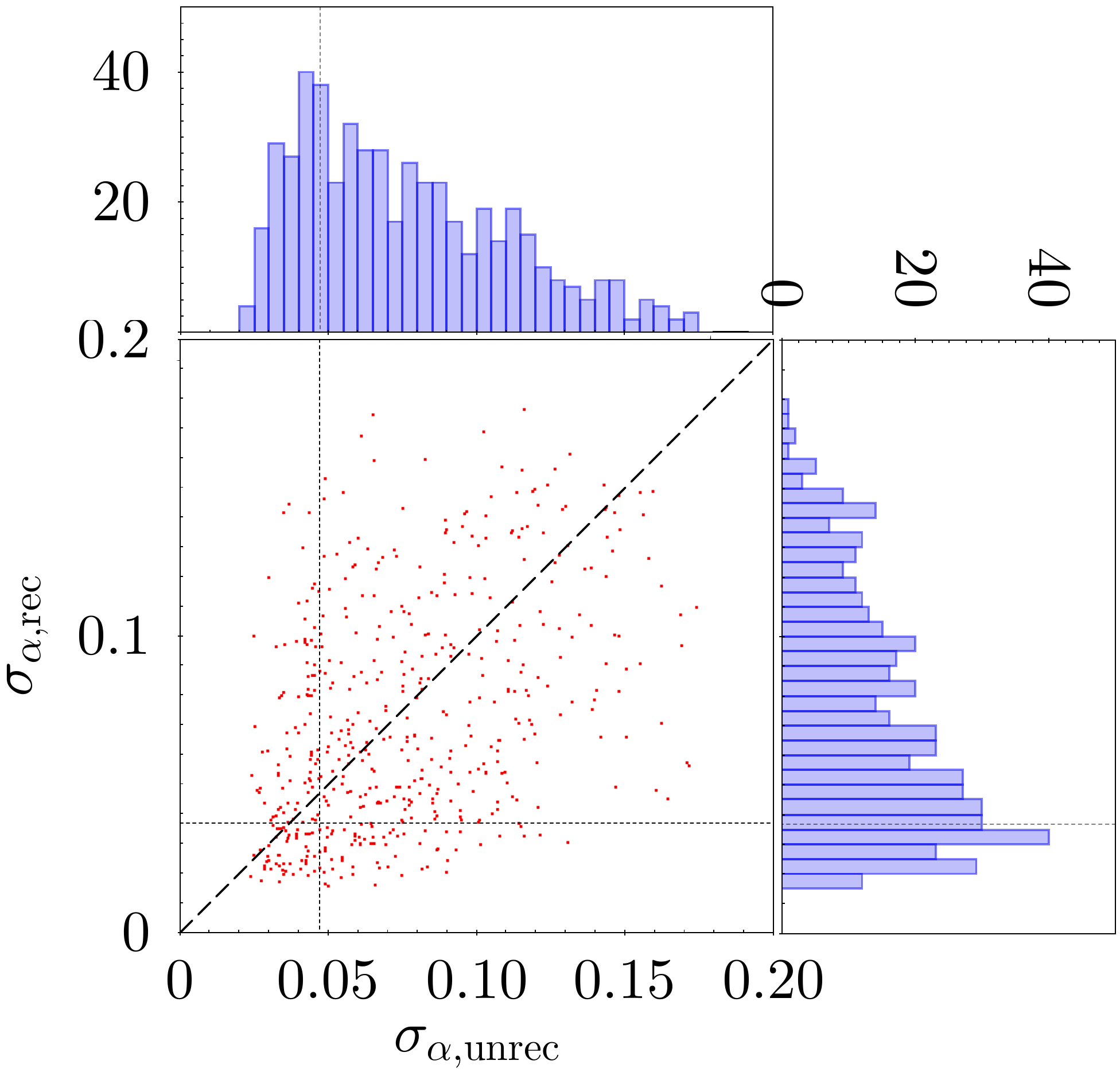}
   \put(-150,125){$39\%$ of points}
   \put(-125,60){$61\%$ of points}}
\caption{As Figure~\ref{fig:histo_rspace} but in redshift-space. The quality of the reconstruction mildly worsen with respect to real-space, however this data analysis is based non-informative flat priors.}
\label{fig:histo_sspace}
\end{figure*}

\begin{table}
	\centering
	\caption{Fit results to average mock correlation functions in redshift-space using eFAM$_{10}$.}
	\label{tab:mean_xi_sspace}
	\begin{tabular}{lcc}
		\hline
		Type & $\alpha$
		 & $\Sigma_\mathrm{NL}\;(h^{-1}\mathrm{Mpc})$\\
		\hline\hline
		pre-recon, $z= 0$ & $1.007 \pm 0.002$ & $11.8\pm0.3$ \\
		post-recon, $z= 0$ & $1.005 \pm 0.002$ & $11.0\pm0.3$ \\
		post-recon, $z=36.6$ & $0.997 \pm 0.001$ & $4.0\pm0.5$ \\
		\hline
	\end{tabular}
\end{table}

The power of the eFAM technique to improve the measurements of the acoustic scale enhancing the BAO signature becomes glaring looking at the monopole of the correlation function, $\tilde{\xi}(s)$ after the non-linear evolution is maximally reversed; see Figure~\ref{fig:eFAM_xi0xi2_med_sspace} top panel. This can be achieved with eFAM at order $M=10$, reaching $z=36.6$ (red line) when the monopole substantially matches the linear model (dashed line). If only the monopole is considered, the eFAM success to correct for RSD at $z=0$ results limited (green line), only slightly improving the measurement of the acoustic scale, moderately sharpening and shifting the BAO peak towards the expected value. The results of the model fitting listed in table~\ref{tab:mean_xi_sspace} show that this reconstruction does not bias the measurements of the acoustic scale and reduces the value of the non-linear broadening $\Sigma_{\mathrm{NL}}$ by 66 percent. The efficiency of the reconstruction in the redshift-space, smaller than in real-space case, could be the result of the lack of precision in the assignment of the initial comoving redshift coordinates, as described in subsection~\ref{subsec:sim}

Although not fully recovering the clustering signal at the BAO scale at $z=0$,  eFAM efficiently restores statistical isotropy already at this redshift, as shown by the vanishing quadrupole of the correlation function after reconstruction; see Figure~\ref{fig:eFAM_xi0xi2_med_sspace}, bottom panel. To ease the comparison at different redshifts pre- and post-reconstruction, the rescaled quadrupole of the two-point correlation function is shown, $\tilde{\xi}_2(s)=5[BD(z)]^{-2}\int_0^1 L_2(\mu)\xi(s,\mu)\mathrm{d}\mu$, with $\mu$ the cosine of the angle between the separation vector and the line-of-sight and $L_2$ the Legendre polynomial of order 2. Before reconstruction, the RSD brake the isotropy of the correlation function returning a non-zero value for $\tilde{\xi}_2$. The deviation from the isotropy is almost completely restored after correcting for the peculiar velocities at the observed redshift $z=0$ and is further improved at small scales when the density field is reconstructed at higher redshift, $z=36.6$.

As done for the real-space analysis, the impact of the reconstruction on the dilation parameter and its error is estimated by a point-wise comparison between the fitted values of $\alpha$ pre- and post-reconstruction from the 512 mocks; see Figure~\ref{fig:histo_sspace}. The distribution of $\alpha_\mathrm{rec}$ is more centred on the actual value, though its dispersions is only mildly improved. The improvement of the precision of $\alpha$ is less significant in redshift-space, here the eFAM algorithm yielding $\sigma_{\alpha,\mathrm{rec}}<\sigma_{\alpha,\mathrm{unrec}}$ for the 61 percent of mocks.

\vspace{-0.3cm}
\section{Conclusions}\label{sec:conclusions}

An extended version of the Fast Action Minimisation method \citep{nusser2000least}, dubbed eFAM, is presented, intended for applications with the next-generation massive spectroscopic surveys designed to observe billions of objects. Based on the \citet{peebles1989tracing} Least Action Principle, the new algorithm coded in C++ reconstructs the trajectories of collisionless mass tracers in generic background cosmologies, owing to a parameterisation of the orbits based on Jacobi polynomials, and works
both in real and redshift-space with a new implementation. It further implements the powerful Poisson solver \textsc{GyrfalcON} \citep{dehnen2002hierarchical}, whose linear scaling with the number of particles realistically allows for application to large catalogues with $O(10^6)$ objects.

For the first time a numerical action method is used for BAO reconstruction. Primarily interested in large scales, where the complexities of galaxy formation and fully non-linear clustering are mitigated, eFAM is probed with mock haloes whose large-scale (quasi-Newtonian) dynamics only mildly deviate from the Hubble flow. A future study will extend this method to mock and real galaxy (rather than halo) catalogues, accounting for the bias and selection function of tracers as done in \citetalias{nusser2000least}.

We have tested the eFAM algorithm on 512 independent halo catalogues extracted from the \textsc{deus-fur} $\Lambda$CDM simulation in real-space, each with about $56,000$ dark matter haloes of mass larger than $1.2\times10^{14}h^{-1}\mathrm{M}_\odot$, typical value for galaxy clusters, in a spherical volume of radius $\sim1h^{-1}$Gpc. Because of the PetaByte-size of the parent catalogue, the peculiar velocities of the FoF haloes were not available. The catalogues in redshift-space are therefore built by modelling the comoving redshift coordinates from the peculiar velocities as reconstructed by eFAM in real-space. Both in real and redshift-space, the reconstructed trajectories are finally trusted only in spheres of radius $\sim700h^{-1}$Mpc, each containing about 23,000 haloes, using the mass in the external shell to model the tidal forces by direct computation; this assures a correct estimation of velocities within a $\sim10$ percent error in redshift-space.

We firstly evaluated the performances of the fully non-linear reconstruction by eFAM, namely using an orbit expansion at 10-th order (eFAM$_{10}$), in recovering the linear model of the monopole of the two-point correlation function in real-space. The eFAM$_{10}$ algorithm successfully recovers the linear correlation function at the BAO scale, reducing the non-linear broadening of the acoustic feature $\Sigma_\mathrm{NL}$ by 87 percent from $9.0\pm0.2h^{-1}$Mpc at $z=0$ to $1.2\pm 0.7h^{-1}$Mpc at $z=6.5$. Moreover, eFAM$_{10}$ returns an unbiased and improved position of the acoustic scale as measured by the dilation parameter, $\alpha_\mathrm{rec}=1.000\pm0.001$, reducing its associated error $\sigma_\alpha$ in 69 percent of the mocks. Instead, the reconstruction achieved by the Zel'dovich approximation obtained from the first-guess, i.e. eFAM$_{1}$, is not equally powerful; the huge errors in the average correlation function post-reconstruction yields a value of $\Sigma_\mathrm{NL}$ larger by a factor $\sim1.7$ and a biased estimation of the BAO scale with 3-10 times larger error. Moreover, this first-order approximation eFAM$_{1}$ only allows for linear trajectories that quickly undergo unphysical crossing, limiting the reconstruction at much lower redshift than eFAM$_{10}$, which more easily removes the effects of non-linear clustering.

Allowing for a reconstruction pushed to very high redshift, the eFAM algorithm is extremely powerful in recovering the BAO feature from anomalous samples that, without reconstruction, would show a wrong location of the BAO peak in monopole of the two-point correlation function (real-space), or no BAO signal at all. Using eFAM$_{10}$ the BAO feature is correctly restored and the signal increased with high statistical significance.

In redshift-space, the fully-non-linear eFAM algorithm successfully corrects for the RSD. By correcting the comoving redshift coordinates of objects using the reconstructed peculiar velocities, eFAM$_{10}$ already restores the isotropy of the two-dimensional correlation function at the observed redshift, here $z=0$. Performing the non-linear reconstruction at the highest redshift possible before shell-crossing, here $z=33.6$, the acoustic ring is efficiently restored. The BAO feature in the azimuthally-averaged two-point correlation function $\xi(s)$ is correspondingly well-sharpened, with a 66 percent reduction of the $\Sigma_\mathrm{NL}$ broadening parameter from $11.8\pm0.3h^{-1}$Mpc to $4.0\pm0.5 h^{-1}$Mpc.
Although not reproducing the internal dynamics in virialised haloes, the fully non-linear eFAM technique achieves a very good accuracy in reconstructing the dynamics down to scales comparable to the mean inter-halo separation, i.e. $\sim10h^{-1}$Mpc, as shown by point-wise comparison of real (simulated) and reconstructed velocities of haloes from small \textsc{deus} simulations. This opens the possibility of a non-parametric modelling of RSD, possibly exploring the effect of local environment on the reconstruction \citep{AchitouvBlake2015}; this is left for a future study.

A final remark on the computational load. The CPU-time is driven by the force computation, which scales linearly with the number of particles $N$, and not by the minimisation procedure. Indeed, the computational efficiency of the code is almost independent of the dimension $M$ of the basis used for the expansion of orbits. The computational complexity increases by a factor of $\sim5$ in redshift-space; eFAM being an iterative reconstruction algorithm, the number of iteration required to relax to a minimum of the action is significantly larger in redshift-space, where the initial conditions of the particles are set by the observed redshifts rather than positions.

\section*{Acknowledgements}

The authors thank A. Sanchez for discussions and helpful suggestions, L. Guzzo, A. Nusser, M. White for fruitful comments, and J.-C. Lambert for the computational support with \textsc{GyrfalcON} that substantially improved the performances of the eFAM code. ES and CS acknowledges partial financial support from LabEx OCEVU  and Action Incitative of Aix-Marseille Universit\'e. EB is supported by MUIR PRIN 2015 ``Cosmology and Fundamental Physics: illuminating the Dark Universe with Euclid'', Agenzia Spaziale Italiana agreement ASI/INAF/I/023/12/0, ASI Grant No. 2016-24-H.0 and INFN project ``INDARK''.




\bibliographystyle{mnras}
\bibliography{BAO_reconstruction_eFAM} 



\appendix

\vspace{-0.5cm}
\section{Basis functions and Jacobi polynomials}\label{appendix:Jacobi}

The Jacobi polynomials $p^{(\alpha,\beta)}_n(x)$, defined for $n=0,1, ... $ and $\alpha,\beta>1$, satisfy the orthogonality condition
\begin{equation}\label{eq:orthogonalityJacobi}
\int_{-1}^1 \mathrm{d} x(1-x)^\alpha (1+x)^\beta p^{(\alpha,\beta)}_n(x) p^{(\alpha,\beta)}_m(x) = h_n\delta_{nm}
\end{equation}
 that can be determined using the recurrence relation
\begin{equation}
p^{(\alpha , \beta)}_{n+1}(x)=(A_nx+B_n)p^{(\alpha , \beta)}_n(x)-C_np^{(\alpha , \beta)}_{n-1}(x).
\end{equation}
For the expression of the coefficients $h_n$, $A_n$, $B_n$, and $C_n$, see \citet{abramowitz1965handbook}.

For $\Lambda$CDM and closer cosmologies the weight function $w(D)= f(D)E(D)Da^2(D)$ in equation~(\ref{eq:orthogonalityBasis}) is almost indistinguishable from a power-law. Defining $x= 2(D/D_\mathrm{obs})-1$, $w(D)$ can be fitted by the weight function $K(1-x)^\alpha(1+x)^\beta=K(2D/D_\mathrm{obs}-2)^\alpha(2D/D_\mathrm{obs})^\beta$ that settles the orthogonality condition~(\ref{eq:orthogonalityJacobi}). The best-fit values for $K$, $\alpha$, and $\beta$ depend on the specific background cosmology; for a standard cold-dark-matter (SCDM) model the exact values $(\alpha,\beta)=(0,3/2)$ are recovered, while for the WMAP-7 $\Lambda$CDM model the best-fit is $(\alpha,\beta)\approx(0,1.53)$.

Once the values of $\alpha$ and $\beta$ are fixed, the functions $q_n^{(\alpha,\beta)}$ for $n\geq1$ are given by
\begin{equation}\label{eq:qn}
q_n^{(\alpha , \beta)}(D) = \int_0^{D_\mathrm{obs}}\mathrm{d}D\,p_n^{(\alpha , \beta)}(D)
=\frac{D_\mathrm{obs}}{n+\alpha +\beta}p_{n+1}^{(\alpha -1, \beta-1)},
\end{equation} 
where the relation
\begin{equation}
\frac{dp_n^{(\alpha , \beta)}}{dx}=\frac{1}{2}(n+\alpha +\beta+1)p_{n-1}^{(\alpha +1 , \beta+1)}
\end{equation}
has been used. The asymptotic limit of the Jacobi polynomials for $x\to -1$, i.e. $D\to 0$, guarantees the vanishing of the initial peculiar velocities while the constrained on the observed positions is satisfy choosing the integration constant in~(\ref{eq:qn}).

\vspace{-0.5cm}
\section{Parametrisation of orbits in redshift-space}\label{appendix:s-space}

The parametrisation of the $i$-th particle's trajectory parallel and perpendicular to the line-of-sight, explicitly accounting for the cosmological dependence, in redshift-space reads
\begin{eqnarray}
 \mathbf{x}^\parallel_i(D) &=&\frac{c\mathbf{s}_{i,0}}{H_0a_0}+\sum_{n=0}^M\mathbf{C}^\parallel_{i,n}Q_n(D)  \label{eq:x_par}\\
 \mathbf{x}^\perp_i(D)&=&\sum_{n=0}^M\mathbf{C}^\perp_{i,n}q_n(D)
\end{eqnarray}
where $Q_n(D)\equiv q_n(D)-(fDE)_\mathrm{obs}p_{n,\mathrm{obs}}$.


\bsp	
\label{lastpage}

\end{document}